\documentclass[11pt]{article}
\usepackage{jcappub,natbib,float,caption,comment,bm,amsmath,wasysym,subfig}
\usepackage{ulem}
\usepackage[usenames,dvipsnames]{xcolor}
\usepackage{hyperref}
\def\be{\begin{equation}}
\def\ee{\end{equation}}
\def\ba#1\ea{\begin{align}#1\end{align}}

\newcommand{\refeq}[1]{equation~(\ref{eq:#1})}

\newcommand{\refEq}[1]{Equation~(\ref{eq:#1})}
\newcommand{\refEqs}[2]{Equations~(\ref{eq:#1})--(\ref{eq:#2})}
\newcommand{\reffig}[1]{figure~\ref{fig:#1}}
\newcommand{\reffigs}[2]{figures~(\ref{fig:#1})--(\ref{fig:#2})}
\newcommand{\refFigs}[2]{Figures~(\ref{fig:#1})--(\ref{fig:#2})}
\newcommand{\refFig}[1]{Figure~\ref{fig:#1}}
\newcommand{\refsec}[1]{section~\ref{sec:#1}}

\newcommand{\dd}{{\rm d}}

\newcommand{\eps}[1]{\epsilon^{#1}}

\def\Mpl{M_{\rm P}}

\title{Tensor Non-Gaussianity from Axion-Gauge-Fields Dynamics: Parameter Search}
\author[a]{Aniket Agrawal,}
\author[b]{Tomohiro Fujita}
\author[a,c]{Eiichiro Komatsu}
\affiliation[a]{Max-Planck-Institut f\"ur Astrophysik, Karl-Schwarzschild-Str. 1, 85741 Garching, Germany}
\affiliation[b]{Department of Physics, Kyoto University, Kyoto 606-8502, Japan}
\affiliation[c]{Kavli Institute for the Physics and Mathematics of the
Universe, Todai Institutes for Advanced Study, the University of Tokyo,
Kashiwa, Japan 277-8583 (Kavli IPMU, WPI)}
\emailAdd{aniket@mpa-garching.mpg.de, t.fujita@tap.scphys.kyoto-u.ac.jp, komatsu@mpa-garching.mpg.de}
\abstract{
We calculate the bispectrum of scale-invariant tensor modes sourced by
spectator SU(2) gauge fields during inflation in a model containing a
scalar inflaton, a pseudoscalar axion and SU(2) gauge fields. A large
bispectrum is generated in this model at tree-level as the gauge
fields contain a tensor degree of freedom, and its production is dominated by
self-coupling of the gauge fields. This is a unique feature of
non-Abelian gauge theory. The shape of the tensor
bispectrum is approximately an equilateral shape for $3\lesssim
m_Q\lesssim 4$, where $m_Q$ is an effective dimensionless mass of the
SU(2) field normalised by the Hubble expansion rate during
inflation. The amplitude of non-Gaussianity of the tensor modes,
characterised by the ratio $B_h/P^2_h$, is inversely proportional to the
energy density fraction of the gauge field. This ratio can be much
greater than unity, whereas the ratio from the vacuum fluctuation of
the metric is of order unity. The bispectrum is effective at
constraining large $m_Q$ regions of the parameter space, whereas the
power spectrum constrains small $m_Q$ regions.
}

\begin{document}

\maketitle
\flushbottom

\section{Introduction}
\label{sec:intro}
We do not yet know how to quantize gravity over an entire
spacetime, but we can quantize its perturbations around a specified
background. In this case, the degrees of freedom of the
gravitational field, including the transverse traceless tensor modes of
the metric, should have ground state vacuum fluctuations
\cite{grishchuk:1974,starobinsky:1979}. These can be found in B-mode
polarisation of the cosmic microwave background (CMB)
\cite{seljak/zaldarriaga:1996,kamionkowski/kosowsky/stebbins:1996};
thus, a detection of non-zero (primordial) B-modes is evidence for
tensor fluctuations of the metric.

So far, no such evidence has been found~\cite{bicep2:2016}. CMB
experiments provide constraints on the tensor-to-scalar ratio $r$, which
is defined as the ratio of the power in tensor modes ($P_h(k_{0})$) to
the power in scalar modes ($P_{\zeta}(k_{0})$), at some wavenumber
$k_0$, $r \equiv P_h(k_{0})/P_{\zeta}(k_{0})$. Currently, this ratio is
constrained to be $r < 0.07$ (95 \% C.L.)~\cite{bicep2:2016} at $k_0 =
0.05$ Mpc$^{-1}$.

We stress here that a detection of B-mode polarisation in the CMB is
evidence for primordial tensor perturbations, but is not necessarily
evidence for the vacuum fluctuation in the tensor metric. For the standard scenario of single-field slow-roll inflation, the tensor fluctuations of the metric, $h_{ij}$, obey the equation
\begin{equation}\label{eq:hij_sourcefree}
\Box h_{ij}(t, \bm x) = 0\,,
\end{equation}
where $\Box$ is the d'Alembertian operator in
4-dimensions. \refEq{hij_sourcefree} shows that if we find evidence for
tensor fluctuations, in the absence of anything that can source them,
they have to be necessarily quantum.

However, there is no {\it a priori} reason to ignore sources in the
right hand side of \refEq{hij_sourcefree}. It is reasonable to think
that there are (many) more than one field during inflation. While their energy
density may be much smaller than that of the dominant inflaton field,
they can still act as sources of perturbations. In general, we write
\begin{equation}\label{eq:hij_sourced}
\Box h_{ij}(t, \bm x) = \Pi_{ij}(t, \bm x)\,,
\end{equation}      
and tensor perturbations are sourced by the anisotropic stress-energy
$\Pi_{ij}$, which is provided by quantum fluctuations of a field other
than the metric. These sourced tensor fluctuations can be much larger
than the vacuum one and can generate observable B-modes, invalidating the claim
that B-modes are evidence for vacuum fluctuations of the
metric. Consequently, there have been intense efforts to build inflation
models where a sizeable $r$ can be generated from sources,
without violating stringent observational constraints on the scalar perturbation. The sources include scalars~\cite{cook/sorbo:2012,carney/etal:2012,biagetti/fasiello/riotto:2013,senatore/silverstein/zaldarriaga:2014}, U(1) gauge
fields~\cite{sorbo:2011,anber/sorbo:2012,barnaby/peloso:2011,barnaby/etal:2012,peloso/sorbo/unal:2016}, and SU(2) gauge
fields~\cite{maleknejad/sheikh-jabbari:2013,dimastrogiovanni/peloso:2012,adshead/etal:2013,adshead/martinec/wyman:2013,maleknejad:2016,dimastrogiovanni/fasiello/fujita:2016,Caldwell:2017chz,Adshead:2016omu}.  

How then, do we differentiate between B-modes generated from vacuum
fluctuations of the metric and those from sources? Vacuum fluctuations of the metric
are usually almost scale invariant, with a slightly red tilt~(see
\cite{kamionkowski/kovetz:2016} for the latest review). On the other
hand, B-modes from sources can have a red or blue tilt or completely
non-power-law spectra such as bumps, depending upon model
parameters. Moreover, the tensor fluctuations produced by sources can be
chiral (see \refsec{amplification}), and so can be seen as a
non-vanishing TB/EB correlation in the
CMB~\cite{Saito:2007kt,namba/etal:2015,Thorne:2017jft,Lue:1998mq,Gluscevic:2010vv,Gerbino:2016mqb}, whereas vacuum
fluctuations produce parity-even B-modes. Finally, these modes can be
highly non-Gaussian \cite{Agrawal:2017awz}. Because tensor modes from
vacuum fluctuations of the metric are almost
Gaussian~\cite{maldacena:2002,maldacena/pimentel:2011}, non-Gaussianity
provides strong evidence for sourced tensor
modes. Therefore, we hope that in any future detections of primordial
gravitational waves (GWs), one would not only check for their amplitude ($r$)
and scale-invariance, but also their non-Gaussianity and
parity-violating correlations. Only after \textit{all} these tests
support nearly scale-invariant, Gaussian, and non-chiral
primordial B-modes, can we confidently claim to have discovered vacuum
fluctuations of the metric.

We focused on a particular set of model parameters in
ref.~\cite{Agrawal:2017awz}, using a model proposed by Dimastrogiovanni,
Fasiello and Fujita \cite{dimastrogiovanni/fasiello/fujita:2016}.
In this paper we shall give more detailed derivations of the bispectrum
and present the results for wider parameter space. The rest of the paper
is organised as follows : in \refsec{model} we present details of the
model that we consider. In \refsec{amplification} we present the second-order Lagrangian for the tensor perturbations in our model and their imprint on the B-mode power spectrum. The third-order Lagrangian is presented in \refsec{bispectrum} and is used to calculate the bispectrum of metric fluctuations. A detailed discussion of the deviation of the bispectrum from the equilateral shape is given in \refsec{peak}. In \refsec{params} we explore parameter regions of the model, which can be potentially observed in upcoming CMB missions. We conclude in \refsec{conclusion}.

\section{Model Setup}
\label{sec:model}
In the model of ref.~\cite{dimastrogiovanni/fasiello/fujita:2016},
inflation is driven by a scalar inflaton $\phi$, which is only minimally
coupled to a pseudoscalar axion $\chi$ and SU(2) gauge fields, $A^a_{\mu}$.
The SU(2) gauge fields and the axion have negligible energy densities
compared to the inflaton, and thus are called ``spectator fields''. They
are coupled to each other by a Chern-Simons like interaction $\chi F \tilde{F}$.  The Lagrangian is then given as,
\begin{align}
\mathcal{L}&=\mathcal{L}_{\rm GR}+\mathcal{L}_{\phi}+\mathcal{L}_{\rm spec},
\label{eq:model_action}
\\
\mathcal{L}_{\rm spec}&=-\frac{1}{2}(\partial \chi)^2-V(\chi)
-\frac{1}{4}F_{\mu\nu}^{a}F^{a\mu\nu}+\frac{\lambda\,\chi}{4f}F_{\mu\nu}^{a}\tilde{F}^{a\mu\nu},
\end{align}
where Einstein gravity $\mathcal{L}_{\rm GR}=\Mpl^2R/2$ is assumed,
the Lagrangian of the inflaton $\mathcal{L}_\phi$ is not specified,
and $\mathcal{L}_{\rm spec}$ denotes the Lagrangian of the spectator fields.
$V(\chi)$ is the potential of the axion field with the canonical kinetic term $-(\partial\chi)^2/2$,
the gauge field strength tensor $F^a_{\mu\nu}$ is written in terms of
the gauge fields as $F^a_{\mu\nu} = \partial_{\mu} A^a_\nu -
\partial_{\nu} A^a_\mu - g \eps{abc}A^b_{\mu}A^c_{\nu}$ with $g$ being
the self-coupling constant, a dimensionless parameter $\lambda$ controls
the strength of the Chern-Simons interaction, $f$ is a decay constant of
the axion field, and $\tilde{F}^{a\mu\nu} \equiv \eps{\mu\nu\rho\sigma} F^a_{\rho\sigma}/(2\sqrt{-g})$ is the dual of $F^a_{\mu\nu}$. In the rest of this section, we discuss the background dynamics, while perturbations will be studied in the following sections.

In this paper, we do not solve for the inflaton $\phi(t)$, but consider
dynamics of the spectator fields in a de Sitter universe,
where the Hubble expansion rate is constant.
We also leave the axion potential $V(\chi)$ unspecified by assuming that
it supports slow-roll of the background axion $\chi_0(t)$ with the
aid of the coupling to the SU(2) fields. While these assumptions are far
from generic, they still capture the essence of physics of generation of
non-Gaussianity, and are observationally
relevant because they produce scale-invariant GWs.

As shown in~\cite{adshead/wyman:2012,Maleknejad:2013npa}, while the
background axion slowly evolves, the homogeneous background component of
the gauge fields has an attractor configuration which respects isotropy of the universe,
\begin{equation}
A^a_0 = 0 \,, \quad A^a_i = \delta^a_i a(t) Q(t),
\label{A BG configuration}
\end{equation}
where $a(t)$ is the scale factor. Then we decompose these spectator
fields into the background and the perturbation components as
\begin{align}
\chi(t,\bm{x}) = \chi_0(t)+\delta \chi(t,\bm{x}),
\qquad
A_i^a(t,\bm{x}) = \delta^a_i a(t) Q(t) +\delta A^a_i(t,\bm{x}).
\end{align}
There also exist non-dynamical components $\delta A_0^a$ that we integrate out.
The equations of motion (EoM) for the background fields are given by
\begin{align}
\ddot{\chi}_0+3H\dot{\chi}_0+\partial_\chi V(\chi_0)
&=-\frac{3g\lambda}{f}Q^2 \left(\dot{Q}+HQ\right),
\label{chi BEoM}
\\
\ddot{Q}+3H\dot{Q} +\left(\dot{H}+2H^2\right)Q 
+2g^2 Q^3 &=\frac{g\lambda}{f} Q^2 \dot{\chi}_0,
\label{Q BEoM}
\end{align}
where the dots denote cosmic time derivatives $\partial_t$
and $H\equiv \dot{a}/a$ is the Hubble expansion rate.
The terms on the right hand side of eq.~\eqref{chi BEoM}
slow down the time evolution of $\chi_0(t)$ in addition to the Hubble friction term $3H\dot{\chi}_0$, because a non-zero background value $Q(t)$ is sustained by energy transfer from the kinetic energy of $\chi_0$ through the coupling. 
Here we introduce two dimensionless parameters;
\begin{equation}
m_Q(t)\equiv \frac{gQ}{H},
\qquad\qquad
\Lambda(t)\equiv \frac{\lambda Q}{f}.
\end{equation}
Here, $m_Q$ is the effective mass of the SU(2) field around its vacuum expectation value (vev) normalized by the Hubble scale, and $\Lambda$ characterizes the coupling strength between 
$\chi_0$ and $Q$. Note that the right hand side of eqs.~\eqref{chi BEoM}
and \eqref{Q BEoM} are proportional to $m_Q \Lambda$.
We consider the slow-roll regime, $m_Q\gtrsim 1$ and $\Lambda\gg 1$,
in which $Q$ is stabilized by its effective mass and $\chi_0$ is significantly
slowed down by the coupling. We can then drop all the terms with time derivatives in the EoMs except for the r.h.s. of eq.~\eqref{Q BEoM} and find~\cite{adshead/wyman:2012}
\begin{align}
m_Q \simeq \left(\frac{- g^2f \partial_\chi V(\chi_0)}{3\lambda H^4} \right)^{\frac{1}{3}},
\\
\xi\equiv \frac{\lambda \dot{\chi}_0}{2fH}
\simeq m_Q+m_Q^{-1}.
\label{xi relation}
\end{align}

The Einstein equations at the background yield
\begin{align}
3\Mpl^2 H^2 &= \rho_\phi +\frac{1}{2}\dot{\chi}_0 +
V(\chi_0) +\frac{3}{2}(\dot{Q}+HQ)^2+\frac{3}{2}g^2Q^4,
\label{Friedmann}
\\
-\frac{\dot{H}}{H^2} &= \epsilon_\phi +\epsilon_\chi +\epsilon_B +\epsilon_E,
\label{Friedmann2}
\end{align}
where $\rho_\phi$ is the energy density of the inflaton and the
slow-roll parameters are defined as $\epsilon_\phi\equiv
-\dot{\rho}_\phi/6\Mpl^2 H^3$, $\epsilon_\chi=\dot{\chi}^2/2\Mpl^2H^2$,
$\epsilon_E\equiv (\dot{Q}+HQ)^2/\Mpl^2H^2,\ {\rm and}~\epsilon_B\equiv
g^2Q^4/\Mpl^2H^2$.
We shall assume that $\rho_\phi$ dominates in eq.~\eqref{Friedmann}.
In the slow-roll regime, $\dot{Q}\ll HQ$, one finds
\begin{equation}
\epsilon_E\simeq \frac{\epsilon_B}{m_Q^2}.
\label{EB relation}
\end{equation}
The background fields appear only through $H, m_Q, \xi, \epsilon_E$ and $\epsilon_B$ in the EoMs for the perturbations.
Using the relationships, eqs.~\eqref{xi relation} and \eqref{EB relation},
one can eliminate $\xi$ and $\epsilon_E$.
Furthermore, in the slow-roll regime, one can disregard the time variations of $H, m_Q$ and $\epsilon_B$ in the leading order approximation.
Therefore we have three relevant background parameters
$H, m_Q$ and $\epsilon_B$ which are approximated to be constant in this paper.%
\footnote{In ref.~\cite{dimastrogiovanni/fasiello/fujita:2016},
the background dynamics is numerically solved, and the perturbations are
also solved with the time varying background quantities, $m_Q(t),
\epsilon_B(t)$ and $H(t)$. They find that $m_Q, \epsilon_B, H\approx const.$
is a very good approximation for a sufficiently strong coupling, especially when one is interested in the range of wavenumbers observable by CMB.}

\section{Amplification of Gravitational Waves}
\label{sec:amplification}
In this section, we study the tensor perturbations at linear level.
Only the tensor perturbations are amplified due to tachyonic instability, while scalar and vector perturbations are not amplified for $m_Q>\sqrt{2}$ in this model~\cite{Dimastrogiovanni:2012ew}. Although the scalar perturbations of $\chi$ and $A_i^a$ are generated from the vacuum fluctuations,  their contribution to the curvature perturbation $\zeta$ is negligible, unless the energy density of $\chi$ becomes comparable to that of the inflaton after inflation~\cite{dimastrogiovanni/fasiello/fujita:2016}.%
\footnote{The scalar perturbations of $\chi$ and $A_i^a$  directly contribute to $\zeta$ through the density perturbation (e.g. $\delta\rho_\chi \simeq \partial_\chi V\delta\chi$). This channel is negligible, for instance, if $\chi$ reaches its potential minimum (i.e. $V(\chi), \partial_\chi V(\chi)\to0$) during inflation. If $\chi_0$ acquires a non-negligible energy fraction after inflation, however, the contribution to $\zeta$ from $\delta\chi$ may be relevant. This implies that the spectator sector can produce $\zeta$ in a way similar to the curvaton mechanism~\cite{lyth/wands:2002,lyth/wands:2003}. We leave this intriguing possibility for future work. } The vector perturbations decay on super-horizon scales in any case.
Therefore in this paper, we assume that the observed curvature perturbation
was produced from the inflaton fluctuation $\delta\phi$ and concentrate
on the tensor perturbations from the gauge fields.

To calculate the power spectrum and bispectrum of GWs we need to expand the action, \refeq{model_action}, up to second
and third order in perturbations, respectively.
We write the tensor perturbations of the metric and the SU(2) gauge
field~\cite{Dimastrogiovanni:2012ew,adshead/etal:2013} as:
\begin{equation}
g_{ij} = -a^2(\delta_{ij}+h_{ij})\,,\qquad 
\delta A^a_i = t_{ai}+\cdots \,,
\end{equation} 
where $\cdots$ represents the scalar and vector perturbations of the SU(2) gauge fields which we neglect in this paper.
We have imposed the transverse and traceless conditions on $h_{ij}$ and $t_{ij}$,
$\delta^{ij}\mathcal{T}_{ij}=\partial_i\mathcal{T}_{ij}=\partial_j\mathcal{T}_{ij}=0\,$ $(\mathcal{T}=h$ and $t$). The inverse metric is given by
$g^{ij} = -a^{-2}(\delta^{ij}-h^{ij}+h^{ik}h^{kj} + \mathcal{O}(h^3))\,.$
For later convenience, we redefine $h_{ij}$ as
\begin{equation}
\psi_{ij} = \frac{1}{2} a\Mpl h_{ij}.
\label{psi def}
\end{equation}
Precisely speaking $t_{ai}$ is not a tensor, since the index $a$ is not
a spatial index but the label of SU(2) gauge. Nonetheless, under the
background configuration of eq.~\eqref{A BG configuration}, $t_{ai}$
transforms as a tensor in practice, because the gauge index $a$ is identified with a spatial index.

Substituting these in \refeq{model_action}, and expanding up to third order, we obtain the Lagrangian of the tensor perturbations as
\begin{equation}
S_{\rm tensor}=\int\dd\tau\dd^3x \sqrt{-g}\left[L_2 + L_3^{(i)} + L_3^{(ii)} +L_3^{(iii)}\right],
\label{expanded action}
\end{equation}
with \cite{obata/soda:2016}
\begin{align}
L_2&=\frac{1}{2}\psi'_{ij}\psi'_{ij}-\frac{1}{2}\partial_k\psi_{ij}\partial_k \psi_{ij}+\frac{1}{\tau^2}\psi_{ij}\psi_{ij}+\frac{1}{2}t'_{ij}t'_{ij} 
-\frac{1}{2}\partial_l t_{ij}\partial_l t_{ij}
+\frac{2m_Q+m_{Q}^{-1}}{\tau}\epsilon^{ijk} t_{il}\partial_j t_{kl}
\notag\\&\quad-\frac{m_Q ^2+1}{\tau^2}t_{ij}t_{ij}+\frac{2\sqrt{\epsilon_B}}{\tau}\left[\frac{1}{m_Q}\psi_{ij}t'_{ij}-\psi_{jm}\epsilon_{aij}\partial_i t_{am} +\frac{m_Q}{\tau}\psi_{ij}t_{ij}\right],
\label{quadratic L}
\end{align}
where $\tau\simeq -1/aH$ is the conformal time, prime denotes  the
derivative with respect to $\tau$ and we neglect terms suppressed by slow-roll
parameters. The cubic Lagrangian $L_3$ will be discussed in the next section.

From the quadratic Lagrangian above, we obtain the following EoMs for GWs $\psi_{ij}(\tau, \bm x)$ and tensor perturbations of the SU(2) gauge field $t_{ij}(\tau, \bm x)$,
\begin{align}\label{eq:psi_t_eom}
&\psi''_{ij}-\partial_k^2\psi_{ij}-\frac{2}{\tau^2}\psi_{ij} = \frac{2\sqrt{\epsilon_B}}{m_Q\tau}t'_{ij}+\frac{2\sqrt{\epsilon_B}}{\tau}\eps{api}\partial_p t_{aj}+\frac{2\sqrt{\epsilon_B}m_Q}{\tau^2}t_{ij}\,,
\\
&t''_{ij}-\partial_k^2 t_{ij}+\frac{2(2m_Q+m_Q^{-1})}{\tau}\eps{lkj}\partial_k t_{il}+\frac{2(m_Q^2+1)}{\tau^2}t_{ij}=\mathcal{O}(\psi_{ij})\,.
\end{align}
Although terms linear in $\psi_{ij}$ also source the gauge field $t_{ij}$, $\psi_{ij}$ is not as substantially amplified as $t_{ij}$~\cite{adshead/etal:2013,adshead/martinec/wyman:2013,maleknejad:2016,dimastrogiovanni/fasiello/fujita:2016}, and so we ignore its contribution as a source for $t_{ij}$. To solve the dynamics of $\psi_{ij}$ and $t_{ij}$, it is useful to
decompose them with the circular polarisation tensors,
\begin{align}
X_{ij}(\tau, \bm{x}) &= \int \frac{{\rm d}^3k}{(2\pi)^3}e^{i\bm k\cdot\bm x}\Big[e^R_{ij}({\bm k})X_{\bm k}^R(\tau)+e^L_{ij}({\bm k})X_{\bm k}^L(\tau)\Big],
\end{align}
where $X=\psi, t$ and the properties of the polarisation tensors are summarized in appendix.~\ref{Polarisation Tensor}. Note that we normalize $e^p_{ij}$
such that $e_{ij}^R(\bm{k}) e_{ij}^R(-\bm{k})=e_{ij}^L(\bm{k})
e_{ij}^L(-\bm{k})=1$.

To proceed further, we quantize $\psi$ and $t$ and expand them in a perturbative series as~\cite{seery:2008}
\begin{equation}\label{eq:pert_series}
\hat{X}^{p}_{\bm k}(\tau) = \hat{X}_1^{p}(\tau, \bm k)+\hat{X}_2^{p}(\tau, \bm k)+\ldots\,.
\end{equation}
The first order components are written as
\begin{align}
\hat{t}_1^{p}(\tau, \bm k)&=T_1^{p}(\tau, k)\, \hat{a}^p_{\bm k}+T^{p*}_1(\tau, k)\, \hat{a}^{p\dag}_{-\bm k},
\\
\hat{\psi}_1^{p}(\tau, \bm k)&=\Psi_1^{p}(\tau, k)\, \hat{a}^p_{\bm k}+\Psi^{p*}_1(\tau, k)\, \hat{a}^{p\dag}_{-\bm k},
\end{align}
with the creation/annihilation operators, $\hat{a}^p_{\bm k}$ and
$\hat{a}^{p\dag}_{\bm k}$, satisfying $[\hat{a}_{\bm k}^p, \hat{a}_{-\bm k'}^{q\dag}]=(2\pi)^3\delta^{pq}\delta(\bm k+\bm k')$. We only consider GWs sourced by the gauge field in this paper, and assign $\hat{\psi}_1$ the same quantum operator as $\hat{t}_1$. The mode functions of $\hat{X}_1^p$ satisfy linearised EoMs and their solutions induce the second order fields $\hat{X}_2^p$ through non-linear terms in the EoMs.

\begin{figure}[tbp]
    \hspace{-2mm}
  \includegraphics[width=70mm]{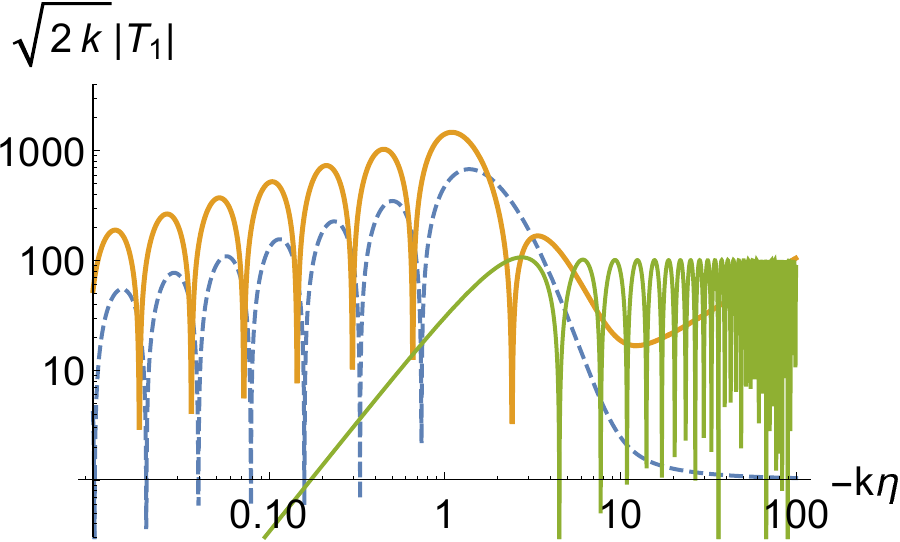}
  \hspace{5mm}
  \includegraphics[width=70mm]{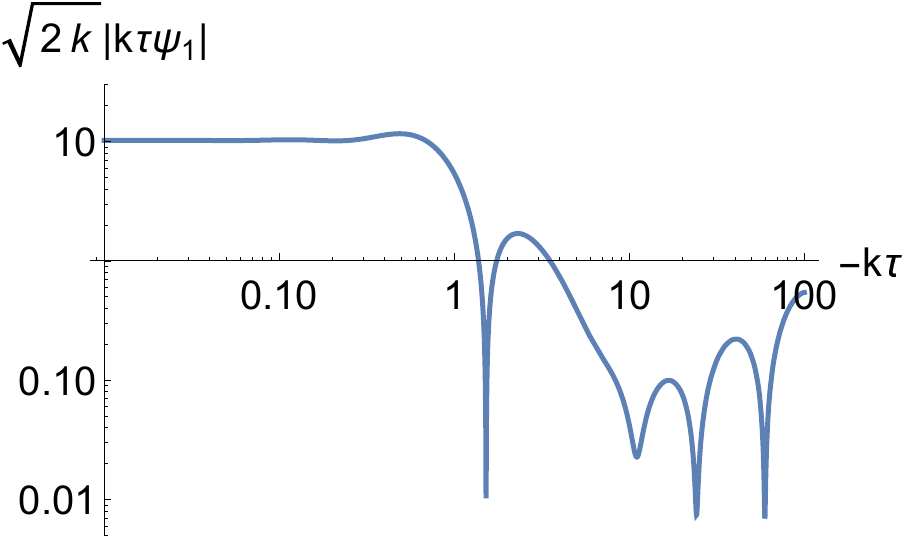}
  \caption
 { {\bf (Left panel)}  We plot the linear gauge tensor mode function $\sqrt{2k}|T_1^R(\eta,k)|$ (blue dashed), the source terms $\sqrt{2k}|m_Q^{-1}\eta \partial_\eta T_1+(m_Q+k\eta)T_1|$ (orange solid) and $10|k\eta \cos(k\eta)-\sin(k\eta)|$ which is proportional to Green's function $G_\psi(\tau,\eta,k)$ in the super-horizon limit, $-k\tau\to 0$, multiplied by $10$ for illustrative purpose (green solid).
 {\bf (Right panel)} The sourced linear gravitational wave $\sqrt{2k}\,|k\tau \Psi_1^R(\tau,k)|$ is shown. 
$|\Psi_1^R|$ grows significantly around the horizon-crossing $(-k\tau\sim 1)$ and stays constant on super-horizon scales.
In both panels, we set $m_Q=3.15$ and $\epsilon_B=3\times 10^{-4}$.}
 \label{Linear}
\end{figure}

In Fourier space the EoMs for the linear mode functions can be written as,
\begin{align}
\label{eq:eom_kspace_eR}
&\partial^2_xT^{R/L}_1+\bigg[1\mp\frac{2(2m_Q+m_Q^{-1})}{x}+\frac{2(m_Q^2+1)}{x^2}\bigg]T^{R/L}_1=\mathcal{O}\left(\Psi_1^{R/L}\right),\\
&\partial^2_x\Psi^{R/L}_1+\bigg[1-\frac{2}{x^2}\bigg]\Psi^{R/L}_1=\frac{2\sqrt{\epsilon_B}}{m_Q x}\partial_xT^{R/L}_1\mp\frac{2\sqrt{\epsilon_B}}{x}T^{R/L}_1+\frac{2\sqrt{\epsilon_B}m_Q}{x^2}T^{R/L}_1,
\end{align}
where $x \equiv -k\tau$.
The minus and plus signs are for right- (R) and left-handed
(L) modes, respectively. Only $T_1^R$ undergoes an instability and it sources only $\Psi_1^R$~\cite{maleknejad:2016,maleknejad/sheikh-jabbari:2013,adshead/martinec/wyman:2013,adshead/wyman:2012},
provided that $m_Q$ is positive. 
Therefore, we only consider the right-handed polarisation in the rest of the paper. 
The homogeneous solution for $T^R_1$ can be analytically calculated and is expressed in terms of the Whittaker function $W_{\beta, \alpha}(z)$ as
\begin{eqnarray}\label{eq:t_homo}
T^R_1(\tau, k) = \frac{1}{\sqrt{2k}}e^{\frac{\pi}{2}(2m_Q+m^{-1}_Q)}W_{\beta, \alpha}(2ik\tau)\,,
\end{eqnarray}
where $\alpha \equiv -i\sqrt{2m^2_Q+7/4}$ and $\beta \equiv -i(2m_Q+m^{-1}_Q)$ \cite{maleknejad/sheikh-jabbari:2013,adshead/etal:2013,dimastrogiovanni/fasiello/fujita:2016}.  
$\Psi^R_1$ can then be calculated using Green's function method,
\begin{equation}\label{eq:psi_gf}
\Psi^R_1(\tau, \bm k) = \int_{-\infty}^{\infty} d\eta\, G_{\psi}(\tau, \eta, k)\mathcal{D}(\eta,k) T_1^R(\eta, k),
\end{equation}
with
\begin{align}
&G_{\psi}(\tau, \eta, k) = \frac{\Theta(\tau-\eta)}{k^3\tau\eta}\Bigg[k(\eta-\tau)\cos\big(k(\tau-\eta)\big)+(1+k^2\tau\eta)\sin\big(k(\tau-\eta)\big)\Bigg]\,,\\
&\mathcal{D}(\eta, k) = \frac{2\sqrt{\epsilon_B}}{m_Q\eta}\partial_{\eta}+\frac{2\sqrt{\epsilon_B}}{\eta^2}(m_Q+k\eta)\,,
\end{align}
where $\Theta(x)$ is the unit Heaviside function of $x$. 
In figure~\ref{Linear}, we plot $T_1^R$, the source term $\mathcal{D}T_1^R$,
Green's function $G_\psi$ in the super-horizon limit and the sourced gravitational wave $\Psi_1^R$.
The time integral of the source term $\mathcal{D}T_1^R$ multiplied by $G_\psi$
yields $\Psi_1^R$.
The source term $\mathcal{D}T_1^R$ peaks around the horizon crossing and Green's function stops oscillating there.
As a result, $\Psi_1^R$ is mainly produced around the horizon crossing as well,
as seen in the right panel.
%
%

\refEq{psi_gf} can also be analytically solved and in the super-horizon limit we obtain,
\begin{equation}\label{eq:psi_suphor}
\lim_{\left| k\tau\right| \to\ 0} \Psi^R_1(\tau, \bm k) = \frac{\sqrt{\epsilon_B}}{\sqrt{2k}k\tau}\mathcal{F}(m_Q)\,,
\end{equation}
from which we obtain the power spectrum of $h$ in the super-horizon limit~\cite{dimastrogiovanni/fasiello/fujita:2016}
\begin{equation}\label{eq:pk_hh}
\frac{k^3}{2\pi^2}P_h^{\rm sourced} = \frac{\epsilon_BH^2}{\pi^2M^2_P}\left| \mathcal{F}(m_Q)\right| ^2\,,
\end{equation}
where $\mathcal{F}$ is a function of $m_Q$, whose exact expression is given in~\cite{dimastrogiovanni/fasiello/fujita:2016} ($\mathcal{F}(m_Q)$
here is $\mathcal{F}_B+\mathcal{F}_E/m_Q$ there). The function $\mathcal{F}(m_Q)$ can be approximated to be an exponential function of $m_Q$, $\left| \mathcal{F}(m_Q)\right| \approx e^{1.8 m_Q}$. Note that eq.~\eqref{eq:pk_hh} is derived under the assumption of $(m_Q, \epsilon_B,H)= const.$
However, as long as the time variations of these background quantities are slow, eq.~\eqref{eq:pk_hh} with $m_Q(t), \epsilon_B(t), H(t)$ at the horizon crossing time $k=a(t)H(t)$ gives a good approximation of $P_h^{\rm sourced}(k)$.
It should be also noted that only the right-handed polarisation modes contribute
to the above $P_h^{\rm sourced}(k)$.

\section{Bispectrum of Gravitational Waves}
\label{sec:bispectrum}

In this section, we calculate the tensor bispectrum of the
right-handed GWs, $B^{RRR}_h$, in the super-horizon limit:
\begin{align} \label{BRRR def}
(2\pi)^3\delta\left(\bm{k}_1+\bm{k}_2+\bm{k}_3\right)B_{h}^{RRR}(k_1,k_2,k_3)
&=\lim_{\tau\to0}\left\langle \hat{h}^R(\tau,\bm k_1) \hat{h}^R(\tau,\bm k_2) \hat{h}^R(\tau,\bm k_3)\right\rangle,
\notag\\
&=\lim_{\tau\to0} \left(\frac{2}{a\Mpl}\right)^3
\left\langle \hat{\psi}^R(\tau, \bm k_1)\hat{\psi}^R(\tau, \bm k_2)\hat{\psi}^R(\tau, \bm k_3)\right\rangle.
\end{align}
The three-point correlator of the right-handed GWs $\hat{\psi}^R= \hat{\psi}^R_1+\hat{\psi}^R_2$
can be written as
\begin{multline}\label{eq:psi_bk}
\left\langle \hat{\psi}^R(\tau, \bm k_1)\hat{\psi}^R(\tau, \bm k_2)\hat{\psi}^R(\tau, \bm k_3)\right\rangle = \left\langle \hat{\psi}_1^R(\tau, \bm k_1)\hat{\psi}_1^R(\tau, \bm k_2)\hat{\psi}_2^R(\tau, \bm k_3)\right\rangle
\\
+\left\langle \hat{\psi}_1^R(\tau, \bm k_1)\hat{\psi}_2^R(\tau, \bm k_2)\hat{\psi}_1^R(\tau, \bm k_3)\right\rangle+\left\langle \hat{\psi}_2^R(\tau, \bm k_1)\hat{\psi}_1^R(\tau, \bm k_2)\hat{\psi}_1^R(\tau, \bm k_3)\right\rangle\,,
\end{multline}
because $\hat{\psi}_1^R$ satisfies Gaussian statistics.
We calculate  $\hat{\psi}_2$ using the second order EoMs for the tensor perturbations which are derived from the cubic Lagrangian.

The cubic tensor Lagrangians introduced in eq.~\eqref{expanded action} are given by
\begin{align}
L_3^{(i)}&=c^{(i)}\Bigg[\epsilon^{abc}t_{ai}t_{bj}\left(\partial_i t_{cj}-\frac{m_Q^2+1}{3m_Q\tau}\epsilon^{ijk}t_{ck}\right)-\frac{ m_Q}{\tau} t_{ij}t_{jl}t_{li}\Bigg],
\\
\nonumber L_3^{(ii)}&=c^{(ii)}\psi_{ij} \bigg[\frac{\tau}{2m_Q}\Big\{ t_{il}' t_{jl}'- \partial_i t_{kl} (\partial_j t_{kl}-2 \partial_k t_{jl})-\partial_k t_{il} \partial_k t_{jl} \Big\} 
\\
&\qquad\qquad\quad-\epsilon^{iab}t_{al}\left( \partial_j t_{bl}-\partial_l t_{bj}\right)-\epsilon^{lab} t_{ai}\partial_l t_{bj}
- \frac{3m_Q}{2\tau} t_{il} t_{jl} \bigg],
\\
L_3^{(iii)}&=c^{(iii)}\psi_{ij}\bigg[\frac{1}{m_Q}
\psi_{jk} t_{ik}'+\epsilon^{ajm}\psi_{lm}\partial_i t_{al}
-\psi_{jk} \epsilon^{akl} \partial_l t_{ai}\bigg],
\end{align}
where we organize terms such that $L_3^{(i)}=\mathcal{O}(t^3),
L_3^{(ii)}=\mathcal{O}(\psi t^2)$ and $L_3^{(iii)}=\mathcal{O}(\psi^2
t)$ and we neglect the $\mathcal{O}(\psi^3)$ terms. The coefficients of the cubic Lagrangians are 
\begin{equation}
c^{(i)}=g=\frac{m_Q^2H}{\sqrt{\epsilon_B}M_{\rm Pl}},\quad  c^{(ii)}=\frac{2m_QH}{M_{\rm Pl}},\quad c^{(iii)}=\frac{4\sqrt{\epsilon_B}H}{M_{\rm Pl}}.
\end{equation}
They satisfy a hierarchical relationship,
\begin{equation}
\frac{c^{(ii)}}{c^{(i)}}=\frac{c^{(iii)}}{c^{(ii)}}=\frac{2\sqrt{\epsilon_B}}{m_Q}\ll1.
\end{equation}
The tree-level contributions from $L_3^{(i)}, L_3^{(ii)}$ and $L_3^{(iii)}$ to the tensor bispectrum are
illustrated as Feynman diagrams in \reffig{Feynman}.
As we see below, the contributions from the three diagrams to the gravitational wave bispectrum are also hierarchical, $(i)>(ii)\gg(iii)$.\footnote{The diagram (ii) includes only two circled crosses which carry a small parameter $\sqrt{\epsilon_B}$,
while the diagram (i) includes three. Hence, in spite of the hierarchical vertex coefficients $c^{(i)}\gg c^{(ii)}$, their contributions are comparable. }
In what follows, we calculate these three contributions in order.
When we show plots in this section, we use the following parameters as an example,
\begin{equation}
H = 8 \times 10^{12} {\rm GeV},\quad m_Q = 3.15,\quad \epsilon_B = 3\times 10^{-4}.
\label{sample parameters}
\end{equation}
The viable parameter space will be explored in section.~\ref{sec:params}.

%
\begin{figure*}
        \centering
        \includegraphics[width=1\textwidth]{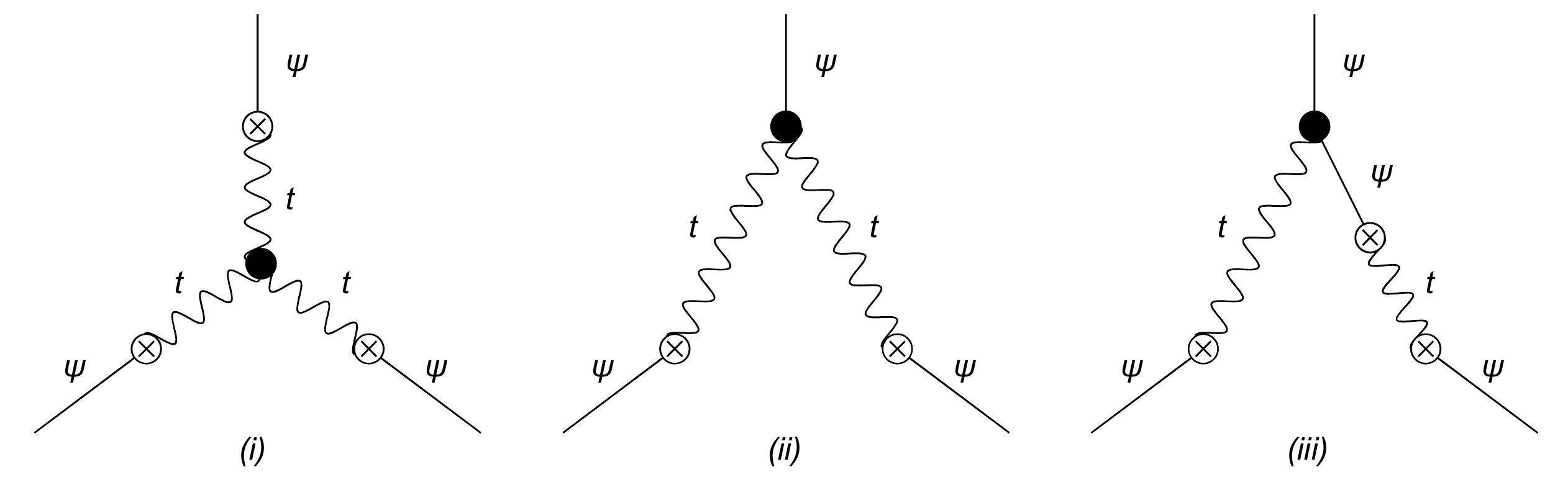}
        \caption{Feynman diagrams illustrating the tree-level contributions
                from the cubic interactions $L_3^{(i)}, L_3^{(ii)}$ and $L_3^{(iii)}$
                to the bispectrum of GWs. The straight and wavy lines show $\psi_{ij}$
                and $t_{ij}$, respectively. The black dots show the vertices of the
                three-point interactions, while the circled crosses show the mixing
                between $\psi_{ij}$ and $t_{ij}$ (the last term in
                Eq.~\eqref{quadratic L}). }
        \label{fig:Feynman}
\end{figure*}

\subsection{Diagram (i)}
\label{sec:bk_d1}

This diagram arises from the self-interaction of the SU(2) gauge field,
and thus is absent in Abelian theory. Here, the second order
gravitational wave $\hat{\psi}_2^R$ is sourced linearly by
$\hat{t}_2^R$, but the second order gauge field perturbation
$\hat{t}_2^R$ is produced by $\hat{t}_1^R$ via non-linearity. The cubic Lagrangian $L_3^{(i)}$ gives the source term $\mathcal{S}^{(i)}_{ij}$
in the EoM for the second order gauge field perturbation,  
\begin{align}\label{eq:t2_t1}
&t''_{ij}(\tau, \bm{x})-\partial_k^2t_{ij}(\tau, \bm{x})+\frac{2(2m_Q+m_Q^{-1})}{\tau}\eps{lkj}\partial_k t_{il}(\tau, \bm{x})+\frac{2(m_Q^2+m_Q^{-1})}{\tau^2}t_{ij}(\tau, \bm{x})=\mathcal{S}^{(i)}_{ij}\,,
\\
&\mathcal{S}^{(i)}_{ij} = \frac{\delta L_3^{(i)}}{\delta t_{ij}} = 2g\eps{aic}t_{al}\partial_l t_{cj}-g\eps{aic}t_{al}\partial_j t_{cl}
-\frac{3gm_Q}{\tau}t_{ip}t_{pj}-\frac{g\xi}{\tau}\eps{ibc}\eps{jmn}t_{bm}t_{cn}\,,
\end{align}
where the source term $\mathcal{S}^{(i)}_{ij}$ is evaluated with the
first order solution $\hat{t}^R_1(\tau, \bm k)$. Although in second
order it is no longer true that the right-handed polarisation is sourced
only by the right-handed tensors, the exponential amplification of the
right-handed modes ensures that terms containing the left-handed modes
are exponentially smaller. Thus we only use the right-handed
polarisation of the gauge field perturbation to evaluate the source
term. Note that this source term contains $g$ explicitly because of the non-Abelian nature of the vertex. 

Expanding the tensor perturbation with the tensor polarisation as before, one finds the EoM in Fourier space as 
\begin{multline} \label{eq:t2_d1_kspace}
\hat{t}_2^{''R}(\tau, \bm k)+\bigg(1+\frac{2(m_Q^2+m_Q^{-1})}{\tau^2}+2k\frac{2m_Q+m_Q^{-1}}{\tau}\bigg)\hat{t}_2^R(\tau, \bm k) = 
\\
g e^{L}_{ij}(\bm{k})\int\int  \frac{d^3q_1}{(2\pi)^3}\,\frac{d^3q_2}{(2\pi)^3}\, \delta_D(\bm{q}_1 +\bm{q}_2 - \bm{k}) Q^{(i)}_{ij}(\bm{q}_1, \bm{q}_2, \tau)\hat{t}^R_1(\bm{q}_1, \tau)\hat{t}^R_1(\bm{q}_2, \tau)\,,
\end{multline}
where we have substituted $[e^{R}_{ij}(\bm k)]^{-1} = e^L_{ij}(\bm k)$ and 
\begin{align}
\nonumber Q^{(i)}_{ij}(\bm{q}_1, \bm{q}_2, \tau) = &\, 2i\eps{aic}e^R_{al}(\bm{q}_1)e^R_{cj}(\bm{q}_2)q_{2l}-i\eps{aic}e^R_{al}(\bm{q}_1)e^R_{cl}(\bm{q}_2)q_{2j}
\\
&-\frac{3m_Q}{\tau}e^R_{ik}(\bm{q}_1)e^R_{kj}(\bm{q}_2)-\frac{m_Q+m_Q^{-1}}{\tau}\eps{ibc}\eps{jmn}e^R_{bm}(\bm{q}_1)e^R_{cn}(\bm{q}_2).
\end{align}
Using the homogeneous solution, \refeq{t_homo}, Green's function for \refeq{t2_d1_kspace} can be written as~\cite{seery:2008},
\begin{align}
G_t(\tau, \eta, k) &= i \Theta(\tau-\eta)\big[T^R_1(\tau, \bm k)T^{*R}_1(\eta, \bm k)-T^{*R}_1(\tau, \bm k)T^R_1(\eta, \bm k)\big],
\\
&= \frac{1}{k}\Theta(\tau-\eta)e^{\pi(2m_Q+m_Q^{-1})}\text{Im}[W_{\beta, \alpha}^*(2ik\tau)W_{\beta, \alpha}(2ik\eta)]\,,
\label{Gt def}
\end{align}
where $\text{Im}(z)$ denotes an imaginary part of a complex number $z$. Dependence of Green's function on the homogeneous solution also ensures that the second order left-handed polarisation of the gauge field is sub-dominant, even if sourced by the first order right-handed polarisation. 

Then, the second order gauge field is given as
\begin{multline}
\hat{t}_2^R(\tau, \bm k) = g e^{L}_{ij}(\bm k) \int_{-\infty}^{\infty} d\eta\,
G_t(\tau, \eta, k) 
\\\times
\int\frac{d^3q_1\,d^3q_2}{(2\pi)^6}\,  \delta_D(\bm{q}_1 +\bm{q}_2 - \bm{k}) Q^{(i)}_{ij}(\bm{q}_1, \bm{q}_2, \eta)\hat{t}^R_1(\eta, \bm{q}_1)\hat{t}^R_1(\eta, \bm{q}_2)\,,
\end{multline}
which yields the second order sourced metric perturbation
(c.f. \refeq{pert_series}) as
\begin{align}
\hat{\psi}^R_2(\tau, \bm k) &= \int_{-\infty}^{\infty} d\eta\, G_{\psi}(\tau, \eta, k)\mathcal{D}(\eta,k) \hat{t}^R_2(\eta, k) \,.
\label{psi2i}
\end{align}
Substituting $\hat{\psi}_1^R$ and the above expression for $\hat{\psi}_2^R$ into eq.~\eqref{eq:psi_bk}, we obtain 
\begin{align}
\nonumber &\left\langle \hat{\psi}_1^R(\tau, \bm k_1)\hat{\psi}_1^R(\tau, \bm k_2)\hat{\psi}_2^R(\tau, \bm k_3)\right\rangle =
\\\notag
&= \int \prod_{i=1}^{3} \Big(d\eta_i G_{\psi}(\tau, \eta_i, k_i)\mathcal{D}(\eta_i,k_i)\Big) \left\langle \hat{t}_1^R(\eta_1, \bm k_1)\hat{t}_1^R(\eta_2, \bm k_2)\hat{t}_2^R(\eta_3, \bm k_3)\right\rangle\,,
\\
\nonumber &= \int \prod_{i=1}^{3} \Big(d\eta_i G_{\psi}(\tau, \eta_i, k_i)\mathcal{D}(\eta_i,k_i)\Big)\, T^R_1(\eta_1, k_1)T^R_1(\eta_2, k_2)g e^{L}_{jl}(\bm k_3)
\int d\eta\,G_t(\eta_3, \eta, k_3) \notag\\ &\times\int\frac{d^3q_1\,d^3q_2}{(2\pi)^6}\, \delta_D(\bm{q}_1 +\bm{q}_2 - \bm{k}_3) Q_{jl}(\bm{q}_1, \bm{q}_2, \eta_3)
 T^{*R}_1(\eta, q_1)T^{*R}_1(\eta, q_2)\left\langle \hat{a}^R_{\bm k_1}\hat{a}^R_{\bm k_2}\hat{a}^{R\dag}_{-\bm q_1}\hat{a}^{R\dag}_{-\bm q_2}\right\rangle\,,
\notag\\
\nonumber &= (2\pi)^3\delta_D(\bm{k}_1 +\bm{k}_2 + \bm{k}_3)\,g\,\Psi^R_1(\tau, k_1)\Psi^R_1(\tau, k_2) \int d\eta_3\, G_{\psi}(\tau, \eta_3, k_3)\mathcal{D}(\eta_3,k_3)  
\\
&\times\int d\eta\, G_t(\eta_3, \eta, k_3)e^{L}_{ij}(\bm k_3)\Big[Q_{ij}(-\bm{k}_1, -\bm{k}_2, \eta)+Q_{ij}(-\bm{k}_2, -\bm{k}_1, \eta)\Big]T^{*R}_1(\eta, k_1)T^{*R}_1(\eta, k_2)\,.
\end{align}
As discussed in appendix.~\ref{Polarisation Tensor}, contraction of the polarisation tensors is calculated as
\begin{equation}
e^{L}_{ij}(\bm k_3)\Big[Q^{(i)}_{ij}(-\bm{k}_1, -\bm{k}_2, \eta)+Q^{(i)}_{ij}(-\bm{k}_2, -\bm{k}_1, \eta)\Big]=-2k_1\Xi\Big[\tilde{\Xi}+(3m_Q+2\xi)/\eta\Big], \end{equation}
where we have defined  
\begin{equation}\label{eq:xi_box}
\tilde{\Xi} = 1+r_2+r_3\,,\quad \Xi = \frac{(1+r_2+r_3)^3}{64r^2_2r^2_3}(r_2+r_3-1)(r_2-r_3+1)(-r_2+r_3+1)\,,
\end{equation}
with $r_2 \equiv k_2/k_1$ and $r_3 = k_3/k_1$. 
Using this, we obtain
\begin{align}
\nonumber &\left\langle \hat{\psi}_1^R(\tau, \bm k_1)\hat{\psi}_1^R(\tau, \bm k_2)\hat{\psi}_2^R(\tau, \bm k_3)\right\rangle = (2\pi)^3\delta_D(\bm{k}_1 +\bm{k}_2 + \bm{k}_3)\,(-2g\Xi k_1)\,\Psi^R_1(\tau, k_1)\Psi^R_1(\tau, k_2) 
\\
&\times\int d\eta_3\, G_{\psi}(\tau, \eta_3, k_3)\mathcal{D}(\eta_3,k_3)  
\int d\eta\, G_t(\eta_3, \eta, k_3)\Big[\tilde{\Xi}+\frac{(5m_Q+2m_{Q}^{-1})}{k_1\eta}\Big]T^{*R}_1(\eta, k_1)T^{*R}_1(\eta, k_2)\,.
\end{align}
Since we are interested in the bispectrum in the super-horizon limit $k\tau \rightarrow 0$, Green's function $G_\psi(\tau,\eta_3,k_3)$ can be reduced. By changing the integration variables from $\eta_3$ and $\eta$ to $y\equiv -k_1\eta_3$ and $z\equiv -k_1\eta$, we obtain
\begin{align}
\nonumber&
\lim_{|k_3\tau|\to0}\int d\eta_3\, G_{\psi}(\tau, \eta_3, k_3)\mathcal{D}(\eta_3,k_3) \int d\eta\, G_t(\eta_3, \eta, k_3)\Big[\tilde{\Xi}+\frac{(3m_Q+2\xi)}{k_1\eta}\Big]T^{*R}_1(\eta, k_1)T^{*R}_1(\eta, k_2) 
\\
\nonumber&=\frac{\sqrt{\epsilon_B}e^{2\pi(2m_Q+m_Q^{-1})}}{\sqrt{r_2}r^4_3k^4_1\tau}\int_{0}^{x_{\text{max}}}\frac{dy}{y^2}\, \Big[r_3y \cos(r_3y)-\sin(r_3y)\Big]
\left(m^{-1}_Q\partial_y+\frac{m_Q}{y}-r_3\right)
\\ \nonumber&\times\int_{y}^{x_{\text{max}}}dz \, \text{Im}\Big[W^*_{\beta, \alpha}(-2ir_3y)W_{\beta, \alpha}(-2ir_3z)\Big]
\left(\tilde{\Xi}-\frac{(5m_Q+2m_Q^{-1})}{z}\right)W^*_{\beta, \alpha}(-2iz)W^*_{\beta, \alpha}(-2ir_2z),
\\ &\equiv \frac{\sqrt{\epsilon_B}e^{2\pi(2m_Q+m_Q^{-1})}}{\sqrt{r_2}r^4_3k^4_1\tau} \mathcal{N}_3,
\end{align}
where we have introduced the UV cutoff $x_{\max}\equiv 2m_Q+m_Q^{-1}+\sqrt{2m_Q^2+2+m_Q^{-2}}$, at which $t_1^R$ starts undergoing a tachyonic instability,
to avoid incorporating unphysical vacuum contributions. The integration
result is not sensitive to the precise value of the cutoff, as we checked numerically for several different values of $x_{\text{max}}$. Using the super horizon solution for $\Psi_1^R$, \refeq{psi_suphor}, and $\psi_{ij} = aM_P h_{ij}/2$ we finally obtain
\begin{align}
\left\langle \hat{h}_1^R(\tau, \bm k_1)\hat{h}_1^R(\tau, \bm k_2)\hat{h}_2^R(\tau, \bm k_3)\right\rangle = (2\pi)^3\delta_D(\bm{k}_1 +\bm{k}_2 + \bm{k}_3) \frac{8g \Xi\epsilon^{3/2}_B}{k^2_1k^2_2k^2_3} \Bigg(\frac{H}{\Mpl}\Bigg)^3 r^{-2}_3\mathcal{F}^2\mathcal{N}_3\,.
\end{align}
The factor of $\delta_D(\bm{k}_1 +\bm{k}_2 + \bm{k}_3)$ ensures the triangle condition, namely that the three wave vectors $\bm k_1$, $\bm k_2$, and $\bm k_3$ form a closed triangle. This is a consequence of homogeneity and isotropy of the Universe. In the same way, the other two terms in eq.~\eqref{eq:psi_bk} can be calculated. Combining them, we obtain the contribution from the diagram (i) as~\cite{Agrawal:2017awz},
\begin{equation}\label{eq:bk_d1}
B_h^{(i)}(k_1, k_2, k_3)=\frac{8m^2_Q \Xi\epsilon_B}{k^2_1k^2_2k^2_3} e^{2\pi(2m_Q+m_Q^{-1})}\Bigg(\frac{H}{\Mpl}\Bigg)^4\Big[\mathcal{F}^{*2}\mathcal{N}_1+ r^{-2}_2|\mathcal{F}|^2\mathcal{N}_2 +r^{-2}_3\mathcal{F}^2\mathcal{N}_3\Big],
\end{equation}
with
\begin{align}\label{eq:tildeN}
&\mathcal{N}_i \equiv \int_0^{x_{\rm max}}\frac{dy}{y^2}\,[r_i y\cos(r_i y)-\sin(r_i y)]
\Big(m^{-1}_Q\partial_y+m_Qy^{-1}-r_i\Big)
\notag\\&\qquad\times \int_{y}^{x_{\rm max}} dz\, {\rm Im}[W^*_{\beta, \alpha}(-2ir_iy)W_{\beta, \alpha}(-2ir_iz)]
 \Big(1+r_2+r_3-\frac{5m_Q+2m_Q^{-1}}{z}\Big)\, \mathcal{W}_i(z),
\end{align}
where   $\mathcal{W}_1(z)=W_{\beta,\alpha}(-2ir_2 z)W_{\beta,\alpha}(-2ir_3z)$,  $\mathcal{W}_2(z)=W^*_{\beta,\alpha}(-2i z)W_{\beta,\alpha}(-2ir_3z)$, and $\mathcal{W}_3(z)=W^*_{\beta,\alpha}(-2ir_2 z)W^*_{\beta,\alpha}(-2iz)$. 

\begin{figure*}
        \centering
        \includegraphics[width=1\textwidth]{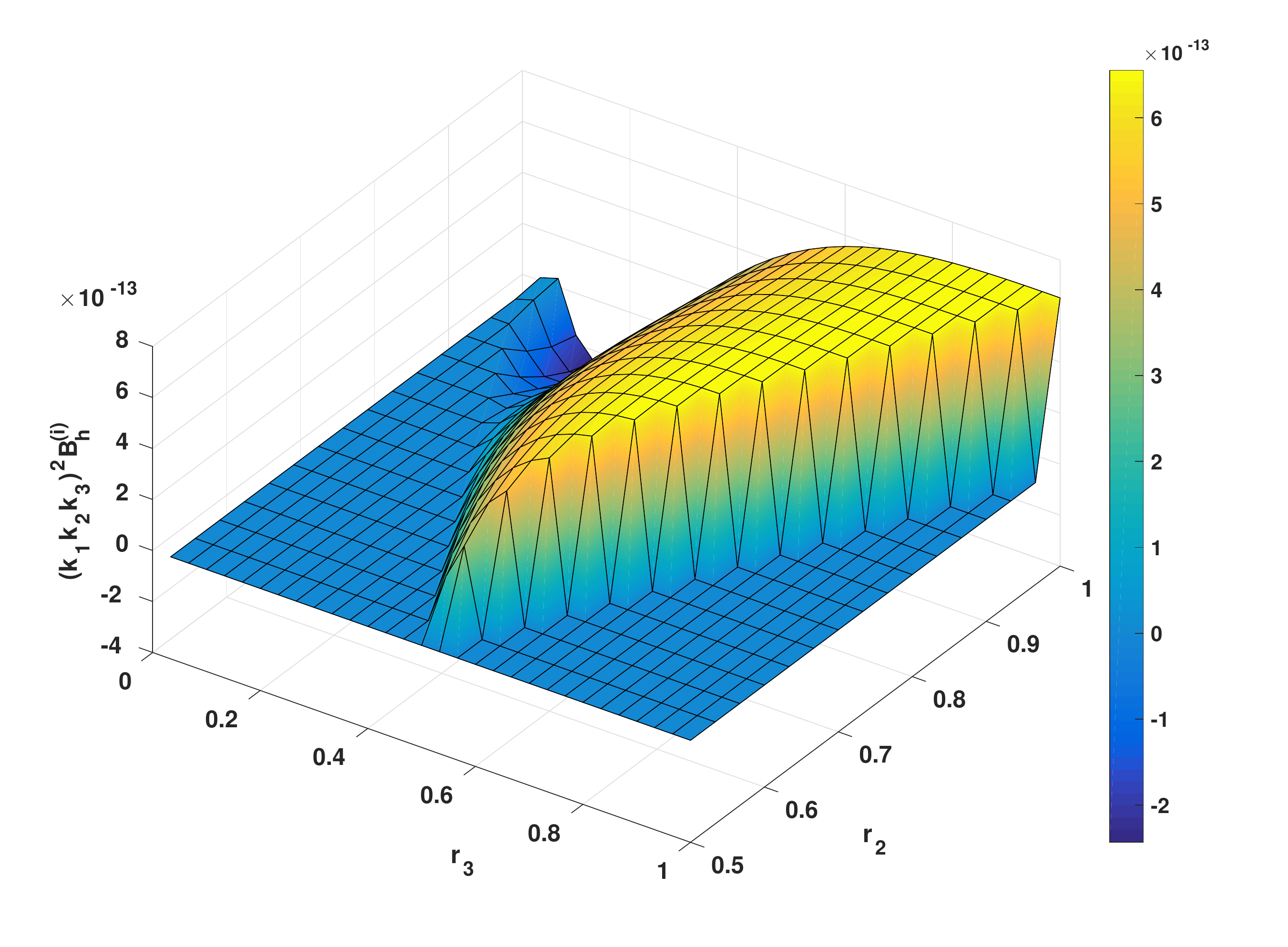}
        \caption{3D plot of $(k_1k_2k_3)^2B^{(i)}_h$ contributed by the
 diagram (i). We show only $r_3 \leq r_2$ and the triangle condition
 implies a bispectrum is non-zero only for $r_2+r_3 \geq 1$. This shape
 has a plateau around $0.6\lesssim r_2\simeq r_3 \le 1$. The parameters
 are  $H = 8 \times 10^{12} {\rm GeV}$,  $m_Q = 3.15$, and $\epsilon_B = 3\times 10^{-4}.$}
        \label{fig:bk_d1_3d}
\end{figure*}

\refFig{bk_d1_3d} shows the tensor bispectrum from the diagram (i). We shall discuss the shape of the bispectrum in detail in section~\ref{sec:peak}. 

\subsection{Diagram (ii)}
\label{sec:bk_d2}
The bispectrum from the diagram (ii) is generated by a non-linear sourcing of the gravitational wave via the $\mathcal{O}(\psi t t)$ terms in the Lagrangian, $L^{(ii)}_3$. The EoM for the second order gravitational wave with the corresponding source terms in Fourier space is given by 
\begin{align}
&\left[\partial_\tau^2+k^2-\frac{2}{\tau^2}\right]\hat{\psi}^{R}_2(\tau, \bm k)=
\frac{H}{\Mpl}[e^{R}_{ij}(\bm{k})]^{-1}S^{(ii)}_{ij}(\tau, \bm k)\,,
\end{align}
where the source term is written as the sum of two parts,
\begin{align}
\mathcal{S}^{(ii)}_{ij}(\tau, \bm k)&\equiv\int\dd^3 x\, e^{-i\bm{k}\cdot\bm{x}}\,\frac{\delta L_3^{(ii)}}{\delta \psi_{ij}(\tau,\bm{x})},
\notag\\&=\int \frac{d^3q_1d^3q_2}{(2\pi)^6}\delta_D(\bm q_1+\bm q_2-\bm k)\left[\mathcal{S}^{(ii)}_{1ij}(\tau, \bm q_1, \bm q_2)+\mathcal{S}^{(ii)}_{2ij}(\tau, \bm q_1, \bm q_2)\right].
\end{align}
$\mathcal{S}^{(ii)}_{1ij}$ has terms without time derivatives,
\begin{align}
\nonumber\mathcal{S}^{(ii)}_{1ij}(\tau, \bm q_1, \bm q_2)&=\Bigg[\frac{3m^2_Q}{\tau}e^R_{ci}(\bm q_1)e^R_{cj}(\bm q_2)-2m_Q\Big\{\eps{acp}e^R_{pj}(\bm q_1)e^R_{ai}(\bm q_2)iq_{2c}
\\
\nonumber&+\eps{apj}e^R_{pc}(\bm q_1)e^R_{ai}(\bm q_2)iq_{2c}-\eps{apj}e^R_{pc}(\bm q_1)e^R_{ac}(\bm q_2)iq_{2i}\Big\}+\tau\Big\lbrace e^R_{ai}(\bm q_1)e^R_{aj}(\bm q_2)q_{1c}q_{2c}
\\
&+e^R_{ac}(\bm q_1)e^R_{ac}(\bm q_2)q_{1i}q_{2j}-2e^R_{ai}(\bm q_1)e^R_{ac}(\bm q_2)q_{1c}q_{2j}\Big\rbrace\Bigg]\hat{t}^R_1(\tau, \bm q_1)\hat{t}_1^R(\tau, \bm q_2) 
\notag\\&\equiv  Q^{(ii)}_{1ij}(\tau, \bm q_1, \bm q_2)\hat{t}^R_1(\tau, \bm q_1)\hat{t}_1^R(\tau, \bm q_2)\,,
\end{align}
whereas $\mathcal{S}^{(ii)}_{2ij}$ comes from time derivatives of $\hat{t}_1^R$,
\begin{align}
\mathcal{S}^{(ii)}_{2ij}(\tau, \bm q_1, \bm q_2) =& \tau e^R_{ai}(\bm q_1)e^R_{aj}(\bm q_2)\hat{t}_1^{'R}(\tau, \bm q_1)\hat{t}_1^{'R}(\tau, \bm q_2)
\notag\\
\equiv&  Q^{(ii)}_{2ij}(\tau, \bm q_1, \bm q_2) \hat{t}_1^{'R}(\tau, \bm q_1)\hat{t}_1^{'R}(\tau, \bm q_2)\,.
\end{align}
The second order gravitational wave is given using Green's function, 
\begin{multline}
\hat{\psi}^R_2(\tau, \bm k) = \frac{H}{\Mpl}\int_{-\infty}^{\infty} d\eta\, G_{\psi}(\tau, \eta, k)\,\int \frac{d^3q_1d^3q_2}{(2\pi)^6}\delta_D(\bm q_1+\bm q_2-\bm k)
\\
\times e^{L}_{ij}(\bm{k})\Big[ Q^{(ii)}_{1ij}(\eta, \bm q_1, \bm q_2)
\hat{t}_1^R(\eta, \bm q_1)\hat{t}_1^R(\eta, \bm q_2)
+Q^{(ii)}_{2ij}(\eta, \bm q_1, \bm q_2)\hat{t}_1^{'R}(\eta, \bm q_1)\hat{t}_1^{'R}(\eta, \bm q_2) \Big]\,.
\end{multline}
We can now compute the first term in eq.~\eqref{eq:psi_bk} produced via the diagram (ii) as
\begin{align}\label{eq:bk_d2}
\nonumber &\left\langle \hat{\psi}_1^R(\tau, \bm k_1)\hat{\psi}_1^R(\tau, \bm k_2)\hat{\psi}_2^R(\tau, \bm k_3)\right\rangle =
\\ \nonumber
& \frac{H}{\Mpl}\int \prod_{i=1}^{2} \Big(\dd\eta_i\, G_{\psi}(\tau, \eta_i, k_i)\mathcal{D}(\eta_i,k_i)\Big)\int \dd \eta_3\, G_{\psi}(\tau, \eta_3, k_3) \int\frac{d^3q_1d^3q_2}{(2\pi)^6}\delta_D(\bm q_1+\bm q_2-\bm k_3)\,
\\ \nonumber
&\qquad\times e^{L}_{jl}(\bm{k})\Big[\left\langle \hat{t}_1^R(\eta_1, \bm k_1)\hat{t}_1^R(\eta_2, \bm k_2)\hat{t}_1^{R}(\eta_3, \bm q_1)\hat{t}_1^{R}(\eta_3, \bm q_2)\right\rangle Q^{(ii)}_{1jl}(\eta_3, \bm q_1, \bm q_2)
\\
&\qquad\qquad\quad+\left\langle \hat{t}_1^R(\eta_1, \bm k_1)\hat{t}_1^R(\eta_2, \bm k_2)\hat{t}_1^{'R}(\eta_3, \bm q_1)\hat{t}_1^{'R}(\eta_3, \bm q_2)\right\rangle Q^{(ii)}_{2jl}(\eta_3, \bm q_1, \bm q_2)\Big]\,.
\end{align}
The expectation values of the 4-point functions are given by\begin{align}
\nonumber
&\left\langle \hat{t}_1^R(\eta_1, \bm k_1)\hat{t}_1^R(\eta_2, \bm k_2)\hat{t}_1^{R}(\eta_3, \bm q_1)\hat{t}_1^{R}(\eta_3, \bm q_2)\right\rangle 
\\
&\qquad= (2\pi)^6 (\delta_{\bm k_1 \bm q_1}\delta_{\bm k_2 \bm q_2}+\delta_{\bm k_1 \bm q_2}\delta_{\bm k_2 \bm q_1})T^R_1(\eta_1, k_1)T^R_1(\eta_2, k_2)T^{*R}_1(\eta_3, q_1)T^{*R}_1(\eta_3, q_2)\,,
\\[5pt] \nonumber 
&\left\langle \hat{t}_1^R(\eta_1, \bm k_1)\hat{t}_1^R(\eta_2, \bm k_2)\hat{t}_1^{'R}(\eta_3, \bm q_1)\hat{t}_1^{'R}(\eta_3, \bm q_2)\right\rangle 
\\
&\qquad= (2\pi)^6 (\delta_{\bm k_1 \bm q_1}\delta_{\bm k_2 \bm q_2}+\delta_{\bm k_1 \bm q_2}\delta_{\bm k_2 \bm q_1})
T^R_1(\eta_1, k_1)T^R_1(\eta_2, k_2)T^{'*R}_1(\eta_3, q_1)T^{'*R}_1(\eta_3, q_2)\,,
\end{align}
where $\delta_{\bm k_1 \bm q_1} \equiv \delta_D(\bm k_1+\bm q_1)$. Note that these functions are invariant under interchange of $q_1\leftrightarrow q_2$. As a result, upon integrating the Dirac delta functions, the polarisation factors in \refeq{bk_d2} yield
\begin{align}
e^{L}_{ij}(\bm{k}_3)\Big[Q^{(ii)}_{1ij}(\eta_3, -\bm k_1, -\bm k_2)+Q^{(ii)}_{1ij}(\eta_3, -\bm k_2, -\bm k_1) \Big] &= 
2\,\Xi \left[\frac{3m^2_Q}{\eta_3}+k^2_1 r_2\eta_3+m_Qk_1(1+r_2)\right]
\,,
\\
e^{L}_{ij}(\bm{k}_3)\Big[Q^{(ii)}_{2ij}(\eta_3, -\bm k_1, -\bm k_2)+Q^{(ii)}_{2ij}(\eta_3, -\bm k_2, -\bm k_1) \Big] &= 2\,\Xi\, \eta_3\,,
\end{align}
where $\Xi\equiv \Xi(r_2, r_3)$ has been defined in \refeq{xi_box}. 
Substituting this in \refeq{bk_d2}, we obtain
\begin{multline}\label{eq:bk_psi_d2}
\left\langle \hat{\psi}_1^R(\tau, \bm k_1)\hat{\psi}_1^R(\tau, \bm k_2)\hat{\psi}_2^R(\tau, \bm k_3)\right\rangle =
(2\pi)^3\delta_D(\bm k_1+\bm k_2+\bm k_3)\, \\
\times 2\,\Xi\,\Psi^R_1(\tau, \bm k_1)\Psi^R_1(\tau, \bm k_2)\frac{H}{\Mpl}\int d\eta_3\, G_{\psi}(\tau, \eta_3, k_3)
\Bigg[\eta_3 T^{'*R}_1(\eta_3, k_1)T^{'*R}_1(\eta_3, k_2)
\\
+\Big\lbrace  r_2 k^2_1\eta_3+m_Qk_1(1+r_2)+\frac{3m^2_Q}{\eta_3}\Big\rbrace T^{*R}_1(\eta_3, k_1)T^{*R}_1(\eta_3, k_2)\Bigg] \,.
\end{multline}
We can similarly evaluate the other two terms in eq.~\eqref{eq:psi_bk} contributed from $L_3^{(ii)}$. Taking the super-horizon limit, we obtain~\cite{Agrawal:2017awz}, 
\begin{align}\label{eq:bk_d22}
k^2_1k^2_2k^2_3 B_h^{(ii)}(k_1, k_2, k_3) = 4\Xi \epsilon_B e^{\pi(2m_Q+m_Q^{-1})}\left(\frac{H}{M_{\rm Pl}}\right)^4
\Big[\mathcal{F}^{*2}\tilde{\mathcal{N}}_1+ r^{-1}_2|\mathcal{F}|^2\tilde{\mathcal{N}}_2 +r^{-1}_3\mathcal{F}^2\tilde{\mathcal{N}}_3\Big],
\end{align}
with
\begin{align}
\tilde{\mathcal{N}}_i \equiv &\int_0^{x_{\rm max}}\frac{dy}{y}\,[r_i y\cos(r_i y)-\sin(r_i y)] \Big[y \tilde{\mathcal{W}}_i(y)
\notag\\& \qquad+\Big(\frac{r_1 r_2 r_3}{r_i}y-(r_1+r_2+r_3-r_i)m_Q+\frac{3m_Q^2}{y}\Big)\, \mathcal{W}_i(y)\Big],
\end{align}
where $\tilde{\mathcal{W}}_1(y)=\partial_y W_{\beta,\alpha}(-2ir_2
y)\partial_y W_{\beta,\alpha}(-2ir_3y)$,
$\tilde{\mathcal{W}}_2(y)=\partial_y W^*_{\beta,\alpha}(-2i y)\partial_y
W_{\beta,\alpha}(-2ir_3y)$, $\tilde{\mathcal{W}}_3(y)=\partial_y
W^*_{\beta,\alpha}(-2ir_2 y) \partial_yW^*_{\beta,\alpha}(-2iy)$, and as before, we have introduced $y \equiv -k_1\tau$.

\begin{figure*}
        \centering
        \includegraphics[width=1\textwidth]{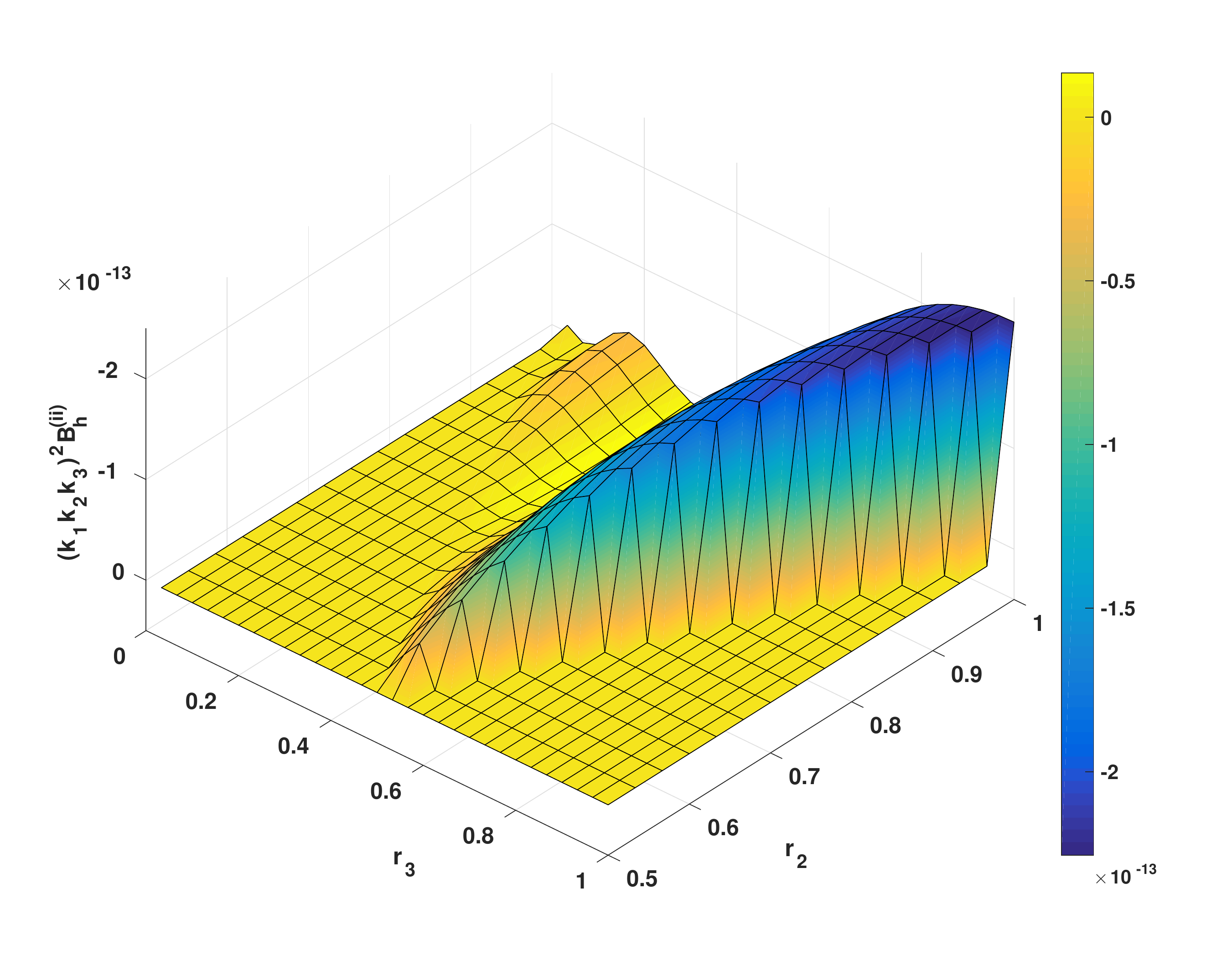}
        \caption{3D plot of $(k_1k_2k_3)^2B^{(ii)}_h$ for diagram
 2. Note that the $z$-axis has been reversed to show negative values. As before, the parameters
 are  $H = 8 \times 10^{12} {\rm GeV}$,  $m_Q = 3.15$, and $\epsilon_B = 3\times 10^{-4}.$}
        \label{fig:bk_d2}
\end{figure*}
\refFig{bk_d2} shows the gravitational wave bispectrum  from the diagram
(ii). The magnitude of this bispectrum is smaller than that of the diagram (i) by a factor of a few for this choice of parameters (however, see \refsec{bk_total} for more details on the relative amplitudes and shape of the bispectrum), and they have the opposite signs. The bispectrum from the diagram (ii) peaks in the equilateral configuration, $r_2 = r_3 = 1$.

\subsection{Diagram (iii)}
\label{sec:bk_d3}
For the diagram (iii), we consider the second order gravitational wave $\hat{\psi}_2^R$ sourced by a first order gauge field perturbation $\hat{t}_1^R$ and a first order metric perturbation $\hat{\psi}_1^R$. The EoM for $\hat{\psi}_2^R$ is derived from $L_{3}^{(iii)}$ to be,
\begin{align}
&
\left[\partial_\tau^2+k^2-\frac{2}{\tau^2}\right]\hat{\psi}^{R}_2(\tau, \bm k)=\frac{H}{\Mpl}[e^{R}_{ij}(\bm{k})]^{-1}\mathcal{S}^{(iii)}_{ij}(\tau, \bm k)\,,
\end{align}
where the source term is again written as the sum of two parts,
\begin{align}
\mathcal{S}^{(iii)}_{ij}(\tau, \bm k)
&\equiv \int\dd^3 x\, e^{-i\bm{k}\cdot\bm{x}}\,\frac{\delta L_3^{(iii)}}{\delta \psi_{ij}(\tau,\bm{x})}, \notag\\
&=
\int \frac{d^3q_1d^3q_2}{(2\pi)^6}\delta_D(\bm q_1+\bm q_2-\bm k)\,4\sqrt{\epsilon_B}\left[\mathcal{S}^{(iii)}_{1ij}(\tau, \bm q_1, \bm q_2)+\mathcal{S}^{(iii)}_{2ij}(\tau, \bm q_1, \bm q_2)\right]\,.
\end{align}
The first part depends on the fields without time derivatives,
\begin{multline}
\mathcal{S}^{(iii)}_{1ij}(\tau, \bm q_1, \bm q_2)=-\hat{t}^R_1(\tau, \bm q_1)\hat{\psi}^R_1(\tau, \bm q_2)\Big[2\left| \bm q_1\right|e^R_{ai}(\bm q_1)e^R_{aj}(\bm q_2)
\\ 
+i \eps{apj}e^R_{ai}(\bm q_1)e^R_{lp}(\bm q_2)q_{1l}+i \eps{ajp}e^R_{al}(\bm q_1)e^R_{lp}(\bm q_2)q_{1i} \Big] \,,
\end{multline}
whereas the second part includes time derivatives of the gauge field perturbation,
\begin{align}
&\mathcal{S}^{(iii)}_{2ij}(\tau, \bm q_1, \bm q_2)= m^{-1}_Q\, e^R_{ai}(\bm q_1)e^R_{aj}(\bm q_2)\,\hat{t}^{'R}_1(\tau, \bm q_1)\hat{\psi}_1^{R}(\tau, \bm q_2) \,.
\end{align}
However, we find that the first part multiplied by the polarisation
factor vanishes: $e^L_{ij}(\bm k_3)\mathcal{S}^{(iii)}_{1ij}(\tau, -\bm
k_1, -\bm k_2) = 0$, for all the permutations of $\bm k_1, \bm k_2$, and
$\bm k_3$. Hence we consider only the second part. With Green's function, we find the second order gravitational wave as
\begin{align}
\nonumber\hat{\psi}^R_2(\tau, \bm k) = \frac{4\sqrt{\epsilon_B}}{m_Q}\frac{H}{\Mpl}\int_{-\infty}^{\infty} &d\eta\, G_{\psi}(\tau, \eta, k)\int
\frac{d^3q_1d^3q_2}{(2\pi)^6}\delta_D(\bm q_1+\bm q_2-\bm k)
\\
&\times\Big[\hat{t}_1^{'R}(\eta, \bm q_1)\hat{\psi}_1^R(\eta, \bm q_2)\,e^{L}_{ij}(\bm{k})e^R_{ia}(\bm q_1)e^R_{aj}(\bm q_2)\Big]\,.
\end{align} 

The first term in eq.~\eqref{eq:psi_bk} from the diagram (iii) yields
\begin{align}\label{eq:bk_d3_raw}
\nonumber &\left\langle \hat{\psi}_1^R(\tau, \bm k_1)\hat{\psi}_1^R(\tau, \bm k_2) \hat{\psi}_2^R(\tau, \bm k_3)\right\rangle =
\frac{4\sqrt{\epsilon_B}}{m_Q}\frac{H}{\Mpl}\int \prod_{i=1}^{2} \left(\dd\eta_i G_{\psi}(\tau, \eta_i, k_i)\mathcal{D}(\eta_i,k_i)\right)\int \dd \eta_3 G_{\psi}(\tau, \eta_3, k_3)
\\
&\times \int\frac{d^3q_1d^3q_2}{(2\pi)^6}\delta_D(\bm q_1+\bm q_2-\bm k_3)\left\langle \hat{t}_1^R(\eta_1, \bm k_1)\hat{t}_1^R(\eta_2, \bm k_2)\hat{t}_1^{'R}(\eta_3, \bm q_1)\hat{\psi}_1^{R}(\eta_3, \bm q_2)\right\rangle e^{L}_{jl}(\bm{k}_3)e^R_{ja}(\bm q_1)e^R_{al}(\bm q_2)\,.
\end{align} 
The expectation value is calculated as
\begin{multline}
\left\langle \hat{t}_1^R(\eta_1, \bm k_1)\hat{t}_1^R(\eta_2, \bm k_2)\hat{t}_1^{'R}(\eta_3, \bm q_1)\hat{\psi}_1^{R}(\eta_3, \bm q_2)\right\rangle= (2\pi)^6 (\delta_{\bm k_1 \bm q_1}\delta_{\bm k_2 \bm q_2}+\delta_{\bm k_1 \bm q_2}\delta_{\bm k_2 \bm q_1})
\\ \times
T^R_1(\eta_1, k_1)T^R_1(\eta_2, k_2)T^{'*R}_1(\eta_3, q_1)\Psi^{*R}_1(\eta_3, q_2)\,.
\end{multline} 
Then, \refeq{bk_d3_raw} reads
\begin{align}\label{eq:bk_psi_d3}
\nonumber &\left\langle \hat{\psi}_1^R(\tau, \bm k_1)\hat{\psi}_1^R(\tau, \bm k_2)\hat{\psi}_2^R(\tau, \bm k_3)\right\rangle = (2\pi)^3\,\delta_D(\bm k_1+\bm k_2+\bm k_3)\,\Psi^R_1(\tau, k_1)\Psi^R_1(\tau, k_2)
\\ 
&\frac{4\sqrt{\epsilon_B}\Xi}{m_Q}\frac{H}{\Mpl}\int d\eta_3\, G_{\psi}(\tau, \eta_3, k_3)
\Big[T^{'*R}_1(\eta_3, k_1)\Psi^{*R}_1(\eta_3, k_2)+T^{'*R}_1(\eta_3, k_2)\Psi^{*R}_1(\eta_3, k_1)\Big]\,.
\end{align} 
Combining with the other two terms in eq.~\eqref{eq:psi_bk}, we obtain  the bispectrum from the diagram (iii) as
\begin{align}\label{eq:bk_d3}
k^2_1k^2_2k^2_3 B_h^{(iii)}(k_1, k_2, k_3) = & 16\Xi \frac{\epsilon^2_B}{m_Q} e^{\pi(2m_Q+m_Q^{-1})}\left(\frac{H}{M_{\rm Pl}}\right)^4
\Big[\mathcal{F}^{*2}\breve{\mathcal{N}}_1+ r^{-1}_2|\mathcal{F}|^2\breve{\mathcal{N}}_2 +r^{-1}_3\mathcal{F}^2\breve{\mathcal{N}}_3\Big],
\end{align}
with
\begin{align}
&\breve{\mathcal{N}}_i \equiv \int_0^{x_{\rm max}}\frac{dy}{y}\,[r_i y\cos(r_i y)-\sin(r_i y)] \breve{\mathcal{W}}_i(y),
\end{align}
where we define $\breve{\mathcal{W}}_1(y)=\partial_y W_{\beta,\alpha}(-2ir_2
y)\Phi(-2ir_3y)+ \partial_y W_{\beta,\alpha}(-2ir_3y)\Phi(-2ir_2y)$, $\breve{\mathcal{W}}_2(y)=\partial_y W^*_{\beta,\alpha}(-2i
y)\Phi(-2ir_3y)+ \partial_y W^*_{\beta,\alpha}(-2ir_3y)\Phi(-2iy)$, $\breve{\mathcal{W}}_3(y)=\partial_y W^*_{\beta,\alpha}(-2iy)\Phi^*(-2ir_2y)+ \partial_y W^*_{\beta,\alpha}(-2ir_2y)\Phi^*(-2iy)$, $y \equiv -k_1\tau$, and
\begin{align}
\nonumber
\Phi(-2i r_i y)\equiv \frac{1}{r_i y}\int_{r_i y}^{x_\text{max}}\frac{dz}{z}
&\left[\left(z-r_iy\right)\cos\left(z-r_iy\right)-\left(1+zr_iy\right)\sin\left(z-r_iy\right)\right]
\\
&\times\left[\frac{\partial_z}{m_Qz}+\frac{m_Q-z}{z^2}\right]W_{\beta,\alpha}(-2iz).
\end{align}

\begin{figure*}
        \centering
        \includegraphics[width=1\textwidth]{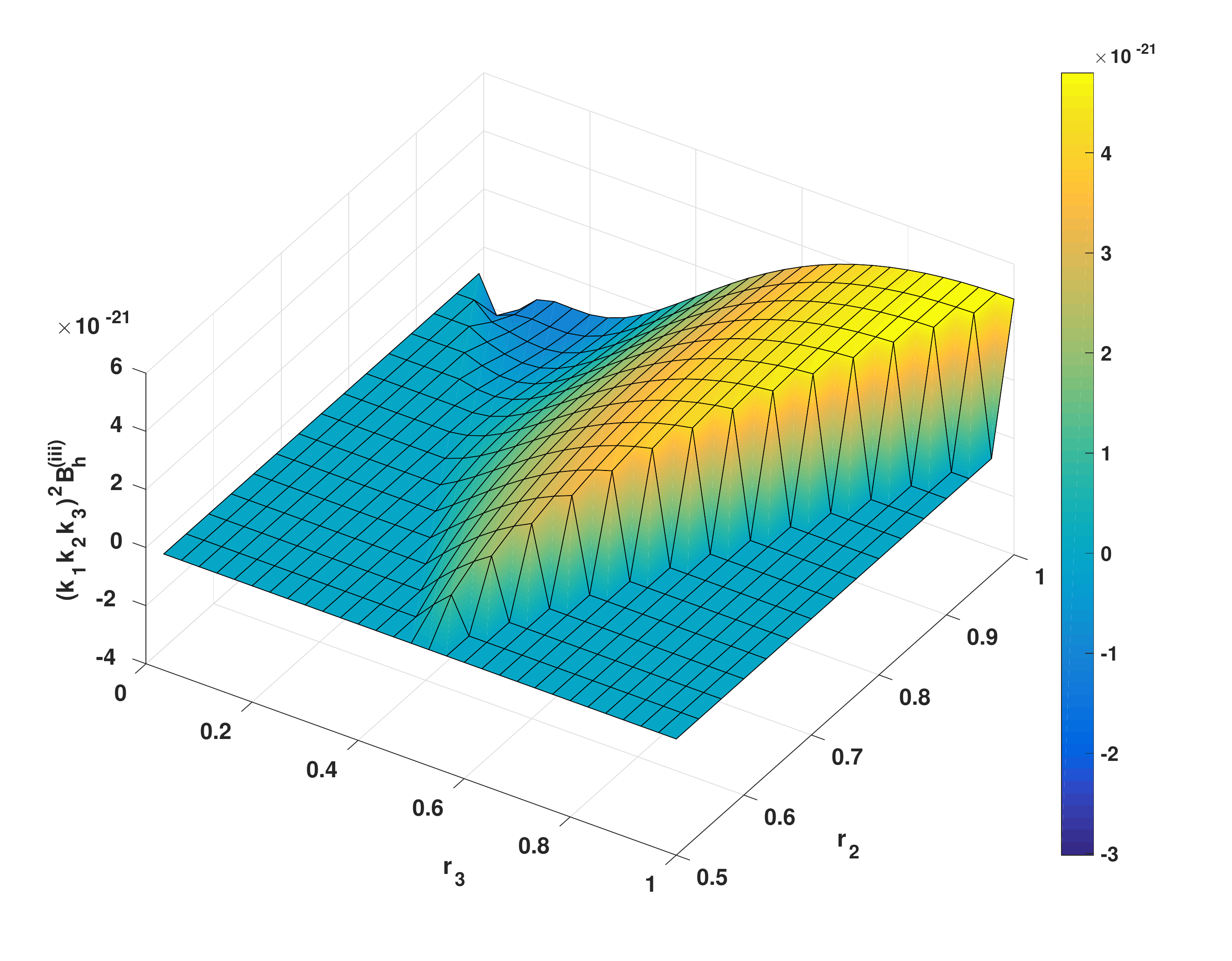}
        \caption{3D plot of $(k_1k_2k_3)^2B^{(iii)}_h$ for diagram 3.
        Its magnitude is much smaller than the other two contributions,
        $B^{(i)}_h$ and $B^{(ii)}_h$, shown in figures \ref{fig:bk_d1_3d}
         and \ref{fig:bk_d2}, respectively. The parameters
         are  $H = 8 \times 10^{12} {\rm GeV}$,  $m_Q = 3.15$, and $\epsilon_B = 3\times 10^{-4}.$}
        \label{fig:bk_d3}
\end{figure*}
\refFig{bk_d3} shows the momentum dependence of the bispectrum from the diagram (iii). We find that the contribution from the diagram (iii) is almost 7 orders of magnitude smaller than that of the diagram (i) and (ii), justifying that we neglected its contribution in our previous work~\cite{Agrawal:2017awz}. This diagram is also zero in the folded limit and the bispectrum  peaks in the equilateral configuration. The contribution from this diagram is so small that we do not compare the templates to it. 

\subsection{Total bispectrum}
\label{sec:bk_total}
\begin{figure*}
        \centering
        \includegraphics[width=1\textwidth]{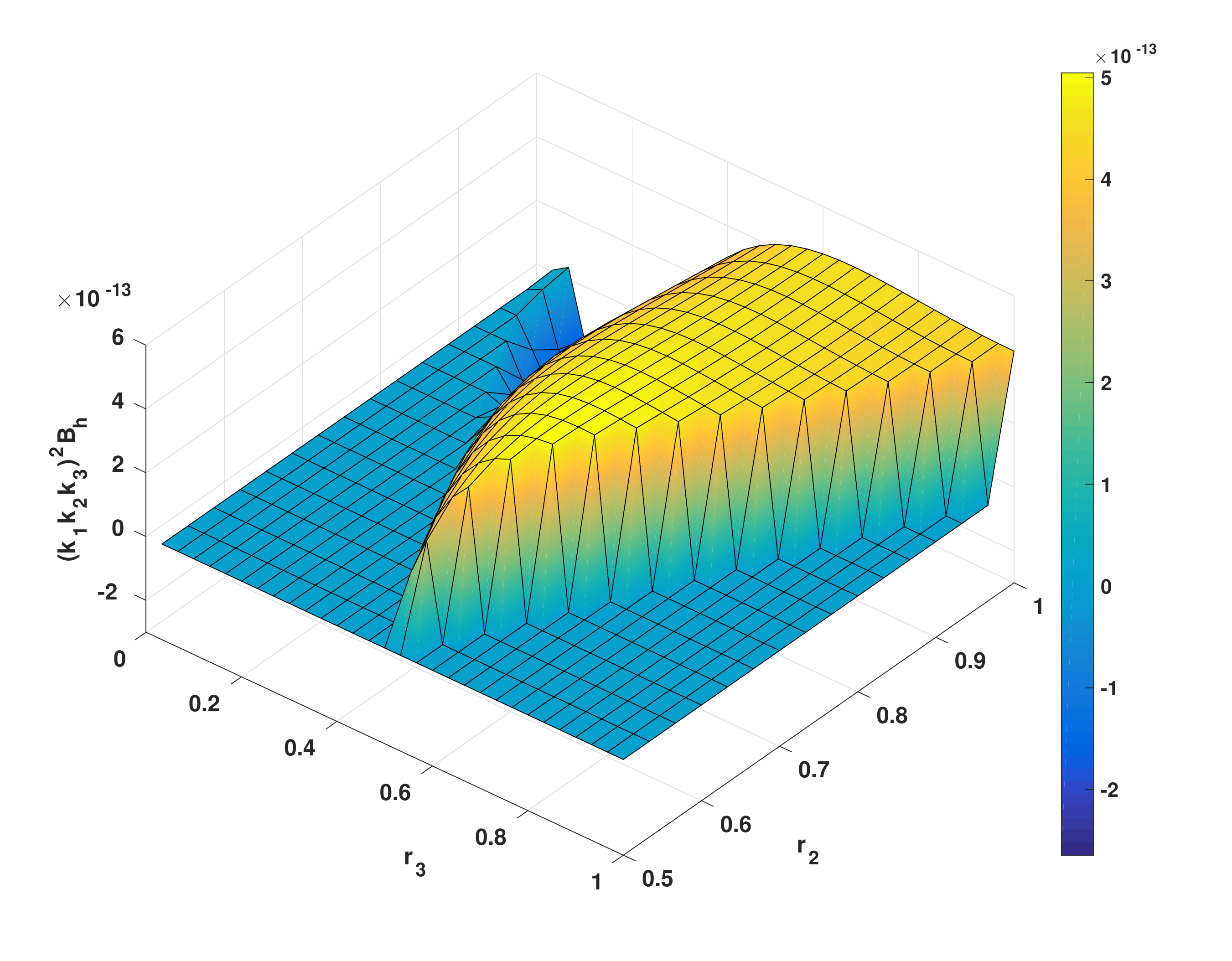}
        \caption{3D plot of the total bispectrum $(k_1k_2k_3)^2B^{RRR}_h$ given in eq.~\eqref{Total Bh}. It has a peak at $r_2\simeq r_3\approx 0.6$
where the brightest yellow is seen. The parameters
are  $H = 8 \times 10^{12} {\rm GeV}$,  $m_Q = 3.15$, and $\epsilon_B = 3\times 10^{-4}.$}
        \label{fig:bk_d5}
\end{figure*}

Combining the three contributions which we have calculated in the previous subsections, eqs.~\eqref{eq:bk_d1}, \eqref{eq:bk_d22} and \eqref{eq:bk_d3}, we obtain the total bispectrum of the sourced GWs in our model as 
\begin{align}\label{Total Bh}
B^{RRR}_h(k_1, k_2, k_3)&=B^{(i)}_h+B^{(ii)}_h+B^{(iii)}_h,
\notag\\
&=\frac{\epsilon_B\,\Xi(r_2,r_3)}{k_1^2k_2^2k_3^2}\left(\frac{H}{\Mpl}\right)^4
\Upsilon(m_Q,r_2,r_3),
\end{align}
with
\begin{align}\label{Upsilon}
\Upsilon(m_Q,r_2,r_3) &\equiv
8m^2_Q e^{2\pi(2m_Q+m_Q^{-1})}\Big[\mathcal{F}^{*2}\mathcal{N}_1+ r^{-2}_2|\mathcal{F}|^2\mathcal{N}_2 +r^{-2}_3\mathcal{F}^2\mathcal{N}_3\Big]
\notag\\&+4e^{\pi(2m_Q+m_Q^{-1})}
\Big[\mathcal{F}^{*2}\tilde{\mathcal{N}}_1+ r^{-1}_2|\mathcal{F}|^2\tilde{\mathcal{N}}_2 +r^{-1}_3\mathcal{F}^2\tilde{\mathcal{N}}_3\Big]
\notag\\&+16 \frac{\epsilon_B}{m_Q} e^{\pi(2m_Q+m_Q^{-1})}
\Big[\mathcal{F}^{*2}\breve{\mathcal{N}}_1+ r^{-1}_2|\mathcal{F}|^2\breve{\mathcal{N}}_2 +r^{-1}_3\mathcal{F}^2\breve{\mathcal{N}}_3\Big].
\end{align}
The $\epsilon_B$ dependence of $\Upsilon$ is weak, since the third line in eq.\eqref{Upsilon} from the diagram (iii) is negligible compared to the others.
In figure~\ref{fig:bk_d5}, we show the shape of the total bispectrum
$(k_1k_2 k_3)^2B^{RRR}_h(k_1, k_2, k_3)$. We find a mild peak around $r_2\simeq r_3 \approx 0.6$
as a result of the combination of the plateau of $B_h^{(i)}$ and 
the negative slope of $B_h^{(ii)}$ on the $r_2=r_3$ plane. 

Aside from some interesting local features, the shape of our bispectrum $B_h^{RRR}$
looks similar to the equilateral shape shown in figure~\ref{fig:eq_shape}.
To quantitatively measure similarity, we calculate the ``cosine''
between the shape of our bispectrum and the equilateral shape,
$\cos(B_h\cdot F_{\rm
eq})$~\cite{babich/creminelli/zaldarriaga:2004}. Definition of the
cosine is described in appendix~\ref{Equilateral Shape}. In
figure~\ref{fig:cosine}, the cosine is shown as a function of
$m_Q$. Note that the cosine depends only on $m_Q$, because $H$ and
$\epsilon_B$ change only the overall amplitude as long as the diagram
(iii) is negligible. We find that the cosine rises from about 0.5 for $m_Q \sim 2$ to around 0.9 for $m_Q \apprge 2.5 $  and it varies up to 1\% for the parameter range of interest (see \refsec{params}). This is because for values of $m_Q < 2.5$ the total bispectrum receives significant negative contribution from the second diagram close to the equilateral limit, thus suppressing the total bispectrum relative to the peak in this region. As a result, the shape becomes quite different from the equilateral shape. 
\begin{figure*}
        \centering
        \includegraphics[width=1\textwidth]{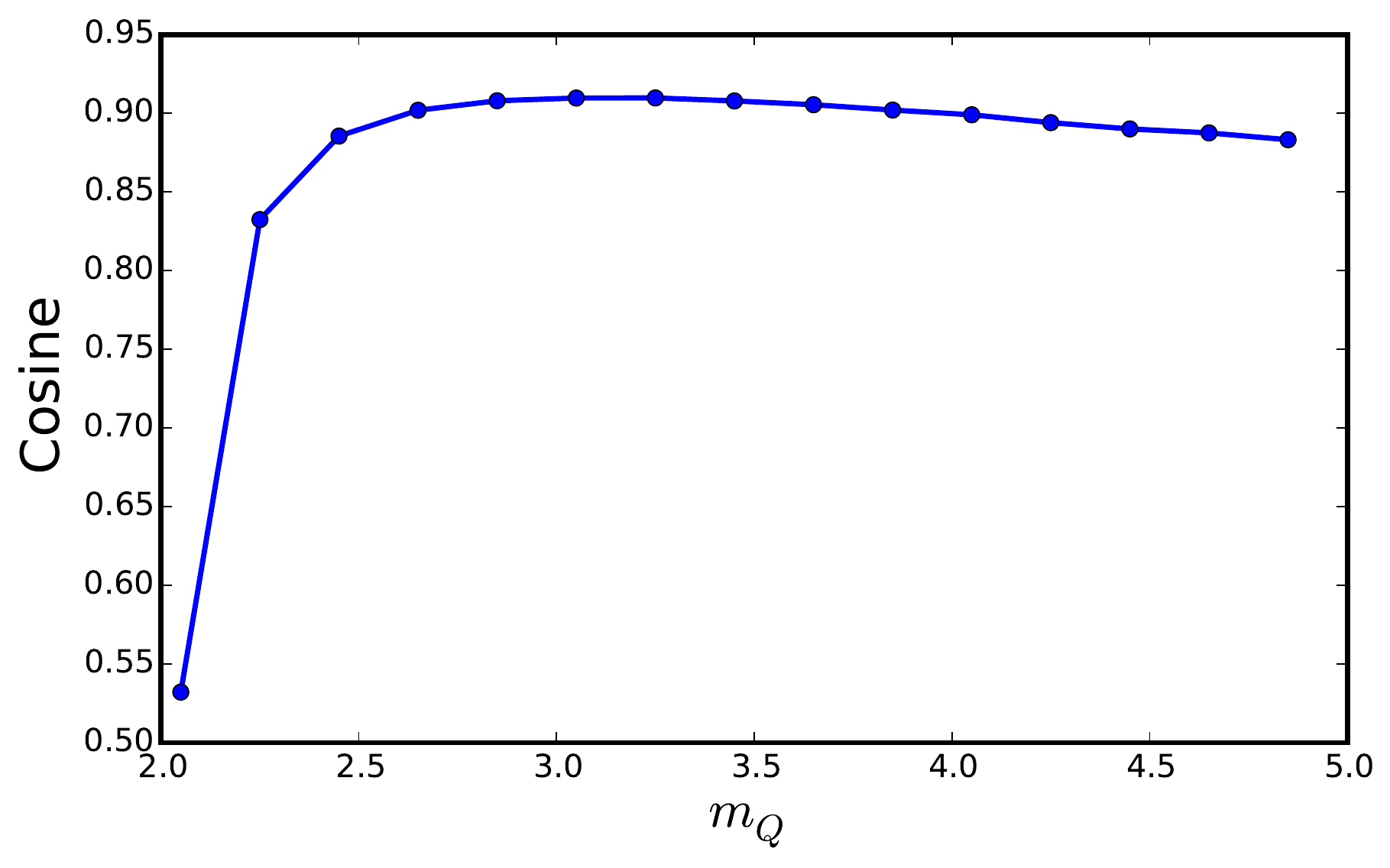}
        \caption{Similarity of the shapes of an equilateral bispectrum and our tensor bispectrum $B_h^{RRR}$, eq.~\eqref{Total Bh}, which is calculated as the ``cosine'' (see appendix~\ref{Equilateral Shape} for definition).}
        \label{fig:cosine}
\end{figure*}

Around 90\% similarity to the equilateral shape implies that our gravitational wave bispectrum is reasonably characterized by the amplitude at the equilateral limit, $r_2=r_3=1$.  
In the equilateral limit, the factors in the bispectrum become
\begin{equation}
\Xi(r_2=r_3=1)=\frac{27}{64} 
\end{equation}
and
\begin{align}\label{eq:smq}
\nonumber\Upsilon_{\rm eq}(m_Q)&\equiv \Upsilon(m_Q,1,1)
\\
 &\simeq  8m^2_Qe^{2\pi(2m_Q+m^{-1}_Q)}\left[ \left| \mathcal{F}\right| ^2 \mathcal{N}_2+2\, \text{Re}[\mathcal{F}^2\mathcal{N}_3]\right]
\notag\\&\quad +4e^{\pi(2m_Q+m^{-1}_Q)}\left[ \left| \mathcal{F}\right| ^2 \tilde{\mathcal{N}_2}+2\, \text{Re}[\mathcal{F}^2\tilde{\mathcal{N}_3}]\right]\,,
\end{align}
where $\text{Re}[z]$ denotes a real part of a complex number $z,$ 
and the small contribution from the diagram (iii) is ignored. 
\begin{figure*}
        \centering
        \includegraphics[width=1\textwidth]{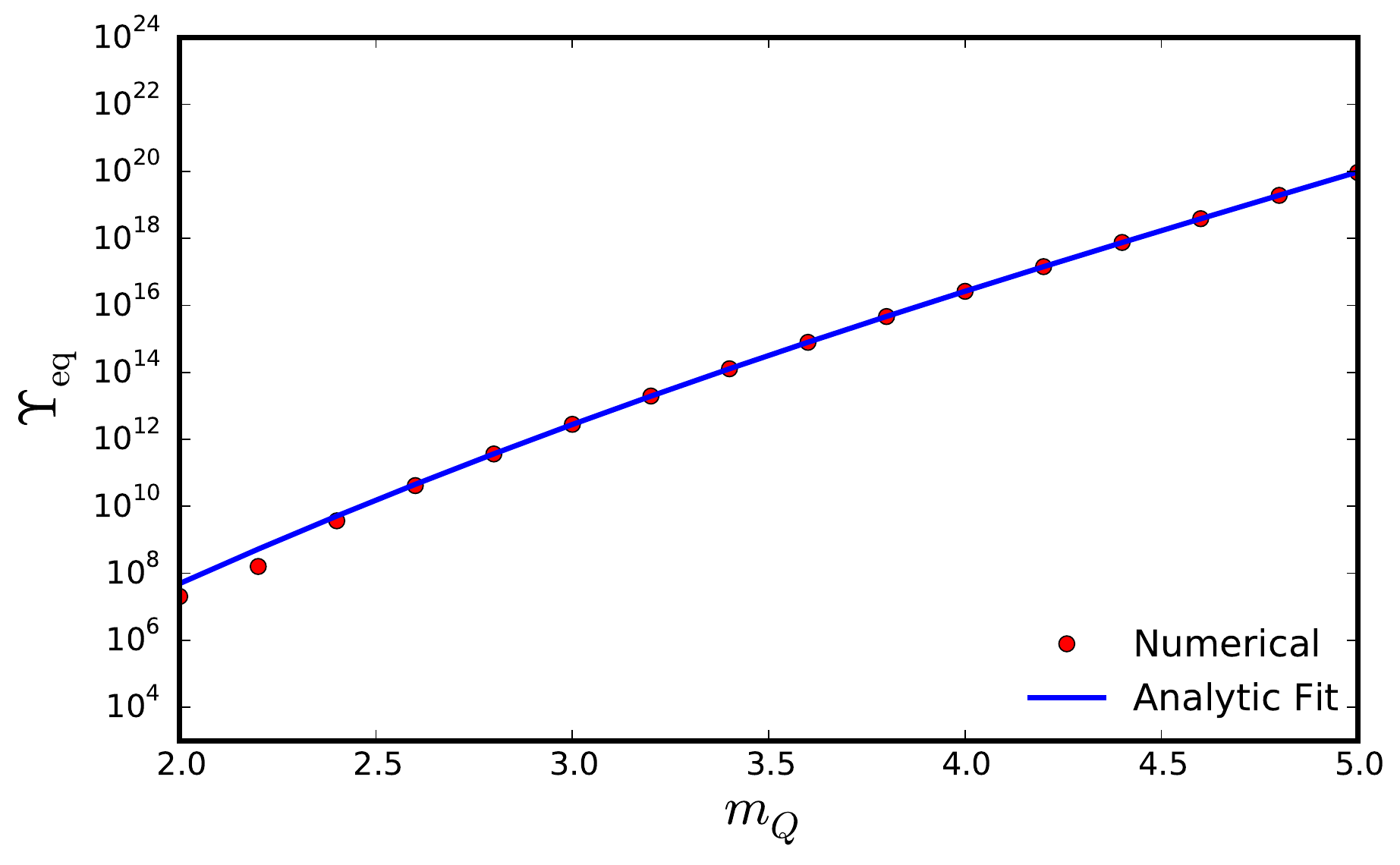}
        \caption{$\Upsilon_{\rm eq}(m_Q)$ defined in eq.~\eqref{eq:smq}
 and its fitting formula (eq.~\eqref{eq:Upsilonfitting}) are plotted as
 the red dots and blue line, respectively.}
        \label{fig:upsilon}
\end{figure*}
In figure~\ref{fig:upsilon}, we plot $\Upsilon_{\rm eq}$. 
For $3\lesssim m_Q\lesssim 5$, $\Upsilon_{\rm eq}$ is well approximated by the following expression: 
\begin{equation}\label{eq:Upsilonfitting}
\Upsilon_{\rm eq}\simeq\exp[0.1377m^3_Q-2.128m^2_Q+18.96m_Q-12.8],
\qquad (2.8 \le m_Q \le 4.8).
\end{equation}
Here, the relative error of this fitting formula is less than $1\%$.

The ratio of the bispectrum to the squared power spectrum of GWs from
the vacuum fluctuation of the metric, $B_h^{\rm vac}/(P_h^{\rm vac})^2$,
is of order unity~\cite{maldacena:2002,maldacena/pimentel:2011}. The
ratio for the sourced GWs can be much greater than unity. From
eqs.~\eqref{eq:pk_hh} and \eqref{Total Bh}, the ratio in the equilateral
limit is given by
\begin{equation}
\frac{B^{\text{sourced}}_h(k,k,k)}{\left(P^{\text{sourced}}_h(k)\right)^2} = \frac{3^3 \Upsilon_{\rm eq}(m_Q)}{2^8 |\mathcal{F}(m_Q)|^4}\,\epsilon_B^{-1}\,.
\end{equation}
\begin{figure*}
        \centering
        \includegraphics[width=1\textwidth]{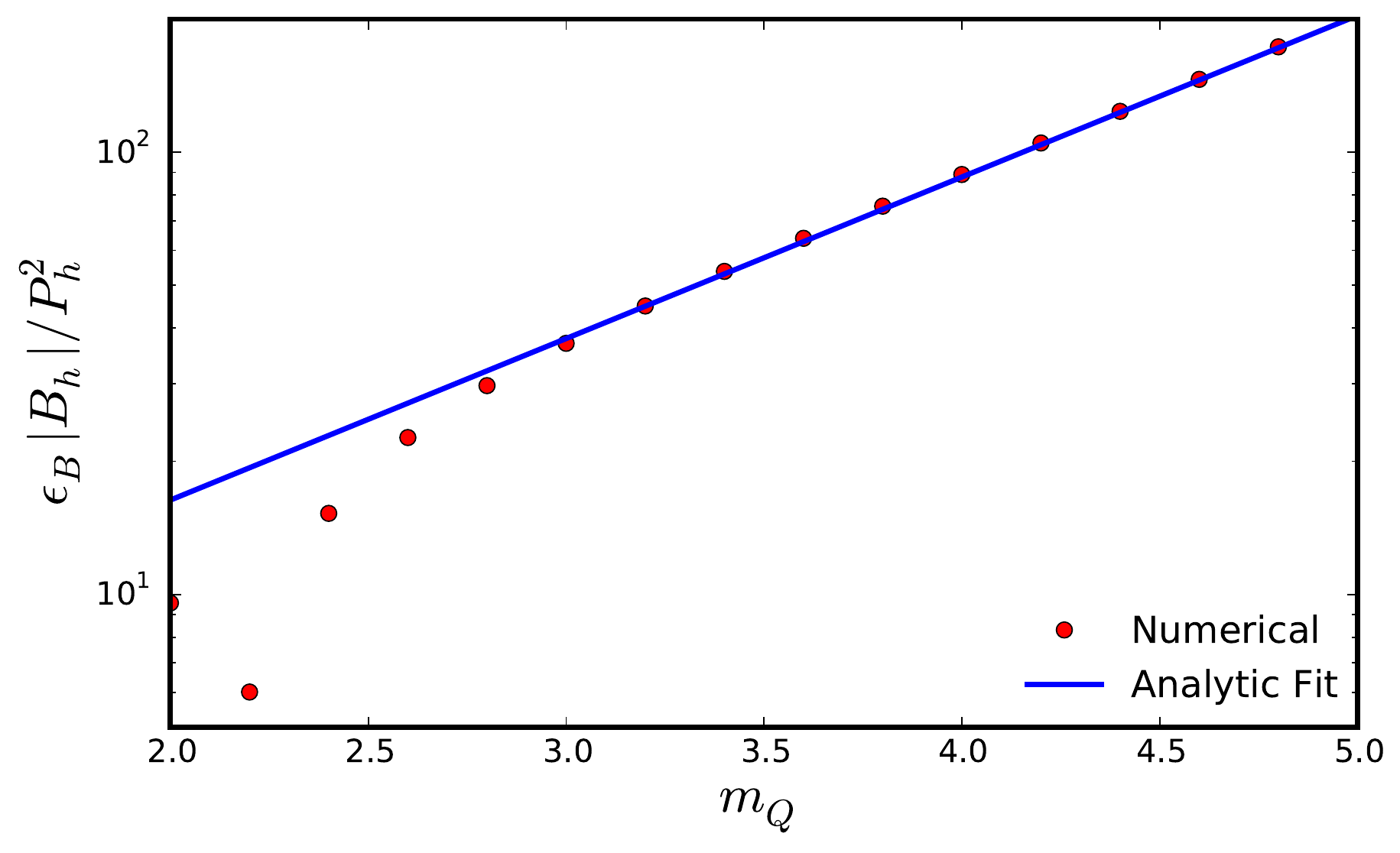}
        \caption{Ratio of the absolute value of the sourced tensor bispectrum to the sourced tensor power spectrum squared (normalised by $\epsilon_B$) as a function of the parameter $m_Q$, along with its fitting formula (eq. \eqref{BP relation}), plotted as the red dots and blue line respectively. It rises exponentially as $m_Q$ increases. Since the $y$-axis shows the ratio normalised by the energy density fraction of the gauge field, $\epsilon_B$, the ratio of the correlation functions can be very large for allowed values of $\epsilon_B$.}
        \label{fig:bhph}
\end{figure*}
In \reffig{bhph}, we plot $\epsilon_B\, B_h^{\rm sourced}/(P_h^{\rm
sourced})^2$ in which steep exponential dependence of the
bispectrum and power spectrum on $m_Q$ cancels out, though milder
exponential dependence remains. We find a simple relation for a specific
range of $m_Q$; for instance, %
\begin{equation}
\frac{B^{\text{sourced}}_h(k,k,k)}{\left(P^{\text{sourced}}_h(k)\right)^2} \approx
\frac{1.816 \, e^{0.841m_Q}}{ \epsilon_B} \simeq \frac{0.908\,e^{0.841m_Q}}{\Omega_A},
\qquad (3\lesssim m_Q \lesssim 5),
\label{BP relation}
\end{equation}
where $\Omega_A\equiv (\epsilon_B+\epsilon_E)/2\simeq (1+m_Q^{-2})\epsilon_B/2\ $ 
(see eq.~\eqref{Friedmann}) is the energy density fraction of the background $SU(2)$ gauge field. Note also, that there is a ``kink" in the bispectrum at $m_Q \sim 2.25$. This corresponds to the value of $m_Q$ for which the bispectrum from the second diagram is larger in magnitude than diagram one at the equilateral configuration, because of which the total bispectrum becomes negative. Since we plot the absolute value of the bispectrum, this appears as a ``kink" in \reffig{bhph}. 

Dependence on the energy density fraction of the gauge fields
in eq.~\eqref{BP relation} is analogous to the curvaton mechanism~\cite{lyth/wands:2002,lyth/wands:2003}, where a similar relation holds for the scalar non-Gaussianity parameter, $f_{\text{NL}}\sim \Omega_\sigma^{-1}$. $\Omega_\sigma$ is the energy density fraction of the curvaton field at its decay time. 
Therefore the origin of the dependence in eq.~\eqref{BP relation} may be
understood in a similar way as the curvaton case~\cite{Komatsu:2003fd}: suppose that the metric perturbation $h$ is given by $h = c_1 t$ where $t$ is the mode function of the gauge field.
At the same time, $t$ is expanded as 
$t = t^{(1)}+t^{(2)}+\mathcal{O}(t^{(3)})$ such that $t^{(2)}=c_2 (t^{(1)})^2$. Then $B_h/P_h^2\sim c_2/c_1$. 
From equations \eqref{psi def} and \eqref{eq:psi_t_eom}, we see that
$c_1 \propto \sqrt{\epsilon_B} H/\Mpl$. We also see from \refeq{t2_t1} that $c_2 \propto g$. 
Thus $B_h/P_h^2 \sim g\Mpl/(\sqrt{\epsilon_B}H)=m_Q^2/\epsilon_B$. 
It should be noted, however, that this relation only holds when the gauge field has the dominant contribution to both the tensor power spectrum and bispectrum, and thus, is not valid in the limit $\epsilon_B \rightarrow 0$.  

\section{Peak of Bispectrum}
\label{sec:peak}
The total tensor bispectrum given in eq.~\eqref{Total Bh} has a peak not
at the equilateral limit $r_2=r_3=1$ but at $r_2=r_3\approx 0.6$ (see figure.~\ref{fig:bk_d5}).
In this section, we study why this happens, by looking into the evolution 
of the tensor perturbations of the $SU(2)$ gauge field $t_{ij}$.

\begin{figure}[tbp]
	\hspace{-2mm}
	\includegraphics[width=70mm]{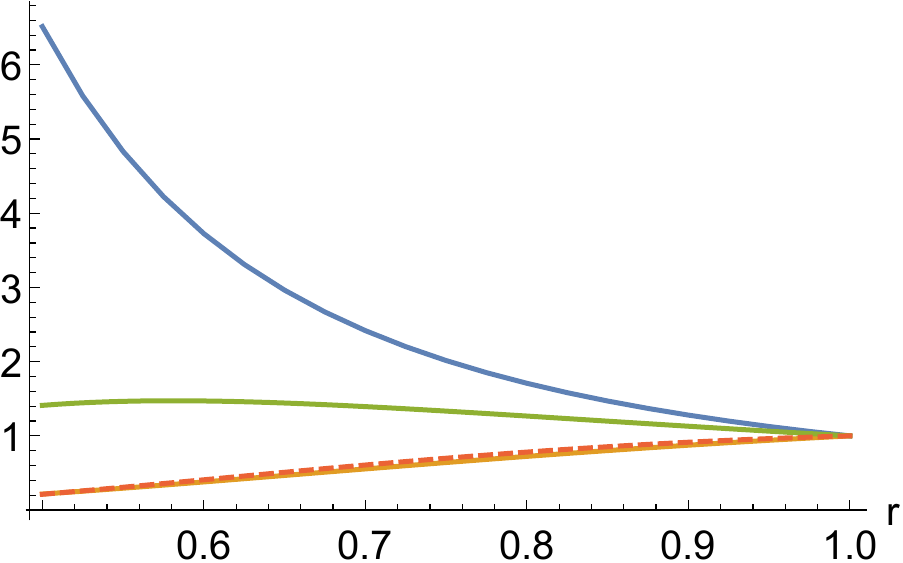}
	\hspace{5mm}
	\includegraphics[width=70mm]{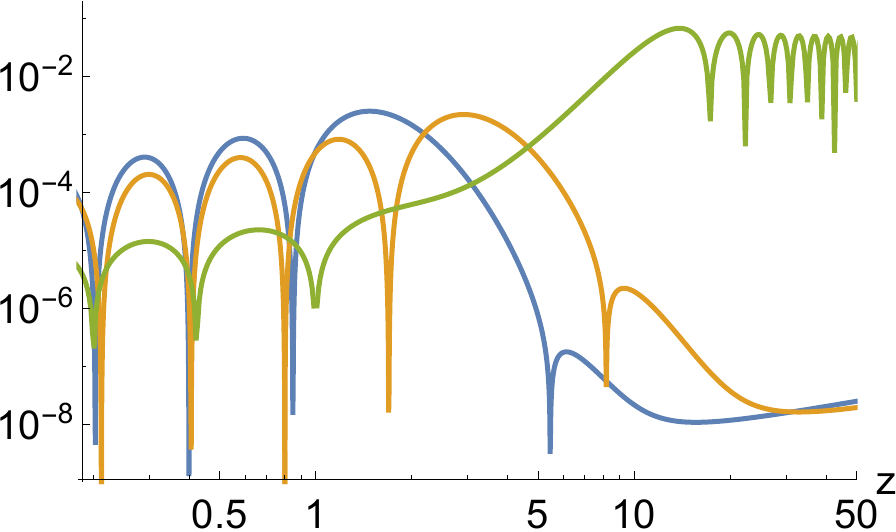}
	\caption
	{{\bf (Left panel)} Absolute values of $\mathcal{N}_i$ and $\tilde{\mathcal{N}}_i$ normalized by their values at $r=1$ are shown, namely $|\mathcal{N}_1(r)|/|\mathcal{N}_1(r=1)|$ (blue), $r^{-2}|\mathcal{N}_2(r)|/|\mathcal{N}_2(r=1)|$ (orange), $|\tilde{\mathcal{N}}_1(r)|/|\tilde{\mathcal{N}}_1(r=1)|$ (green) and $r^{-1}|\tilde{\mathcal{N}}_2(r)|/|\tilde{\mathcal{N}}_2(r=1)|$ (red dashed). 
		{\bf (Right panel)} {Absolute value of the source term part in the integrand of $\mathcal{I}_{1}$, namely $|(z(1+2r)-5m_Q-2m_Q^{-1})\mathcal{W}_1(z)|$,
			is shown for $r=1$ (blue) and $r=0.5$
 (orange). Its Green's function part, $|{\rm Im}[W^*_{\beta,
 \alpha}(-2iy)W_{\beta, \alpha}(-2iz)]|$, for $y=1$ is also plotted as
 the green line.}
		In both panels we set $m_Q=3.15$.}
	\label{Nir}
\end{figure}

The shape of the tensor bispectrum is determined by
$\mathcal{N}_i(m_Q, r_2, r_3)$ and $\tilde{\mathcal{N}}_i(m_Q, r_2,
r_3)\ (i=1,2,3)$ as well as $\Xi(r_2,r_3)$ in eq.~\eqref{Total Bh}, while the contributions from $\breve{\mathcal{N}}_i(m_Q, r_2, r_3)$ are negligible.
For our current purpose, it suffices to focus on the case with $r_2=r_3$
in which the three momenta, $\bm{k}_1,\bm{k}_2,\bm{k}_3,$ form an obtuse-angled isosceles triangle. In other words, we concentrate on a cross-section surface
of the 3D plot, figure~\ref{fig:bk_d5}.
For $r\equiv r_2=r_3$, the $r$ dependence of  $\mathcal{N}_i(m_Q, r)$ and $\tilde{\mathcal{N}}_i(m_Q, r)$ is shown in figure~\ref{Nir}. 
We do not plot $|\mathcal{N}_3|$ and $|\tilde{\mathcal{N}}_3|$
which are the same as $|\mathcal{N}_2|$ and $|\tilde{\mathcal{N}}_2|$
for $r_2=r_3$, respectively.
We find that only $\mathcal{N}_1$ grows significantly as $r$ decreases,
while the others have moderate dependences.
To understand its behaviour, we look closely at the second line
of $\mathcal{N}_1$ in eq.~\eqref{eq:tildeN} for $r_2=r_3$,
\begin{equation}
\mathcal{I}_1(m_Q,r,y)\equiv\int_{y}^{x_{\rm max}} \frac{\dd z}{z}\, {\rm Im}[W^*_{\beta, \alpha}(-2iy)W_{\beta, \alpha}(-2iz)]
 \Big(z(1+2r)-5m_Q-2m_Q^{-1}\Big)\, \mathcal{W}_1(z).
 \label{I1 def}
\end{equation}
{In the integrand of $\mathcal{I}_1$,} the first part, ${\rm Im}[W^*_{\beta, \alpha}(-2iy)W_{\beta, \alpha}(-2iz)]$, represents Green's function for $t_2^R$ given in eq.~\eqref{Gt def}, and the second part, $\left(z(1+2r)-5m_Q-2m_Q^{-1}\right)\, \mathcal{W}_1(z)$, represents the non-linear source term from the first order tensor perturbations, $t_1^R\times t_1^R$.
They are plotted in figure~\ref{Nir} for $r=1$ and $r=0.5$.
Basically the second part is shifted by a factor of $\approx 2$ along the $z$-axis, as $r$ is reduced to the half. 
{However, Green's function has a bigger amplitude at larger $z$ without oscillations up to $z\simeq 10$.
	Note that the non-linear source term contains
	$\mathcal{W}_1(z)\equiv W_{\beta,\alpha}(-2ir_2
	z)W_{\beta,\alpha}(-2ir_3z)$, indicating that the two sourcing modes $t_1^R$ have momenta $k_2=k_3=r k_1$
	in the case of $r\equiv r_2=r_3$, while the momentum of the sourced mode $t_2^R$ is $k_1$ in the process of $\mathcal{N}_1$.
	The fact that Green's function for $t_2^R$ is larger on sub-horizon scales implies that the source effect from $t_1^R\times t_1^R$ to $t_2^R$ is more efficient when the sourcing modes $t_1^R(rk_1)$ have lower momenta (i.e. a smaller $r$) and get amplified before $t_2^R(k_1)$ crosses the horizon. In other words,  $\mathcal{N}_1$ becomes larger for a smaller $r$, because atypical Green's function $G_t$ allows the sourcing effect to be active deep inside the horizon.}
%

This non-linear sourcing process of $t_2$ through the diagram (i) shows clear contrast from the linear sourcing process from $t_1$ to $\psi_1$ discussed in section~\ref{sec:amplification}.
There, Green's function for $\psi$, $G_\psi$, rapidly oscillates inside the horizon
and does not allow $t_1$ to induce $\psi_1$ on sub-horizon scales, as
shown in figure~\ref{Linear}. In cases where only such normal Green's
functions are involved, the shape of the bispectrum is typically equilateral, since all the modes are mainly produced
around the horizon crossing. Nonetheless, in our case, Green's function for $t_R$ is peculiar
due to tachyonic instability, and the peak of the bispectrum deviates from the equilateral limit.

\begin{figure}[tbp]
\begin{center}
  \includegraphics[width=100mm]{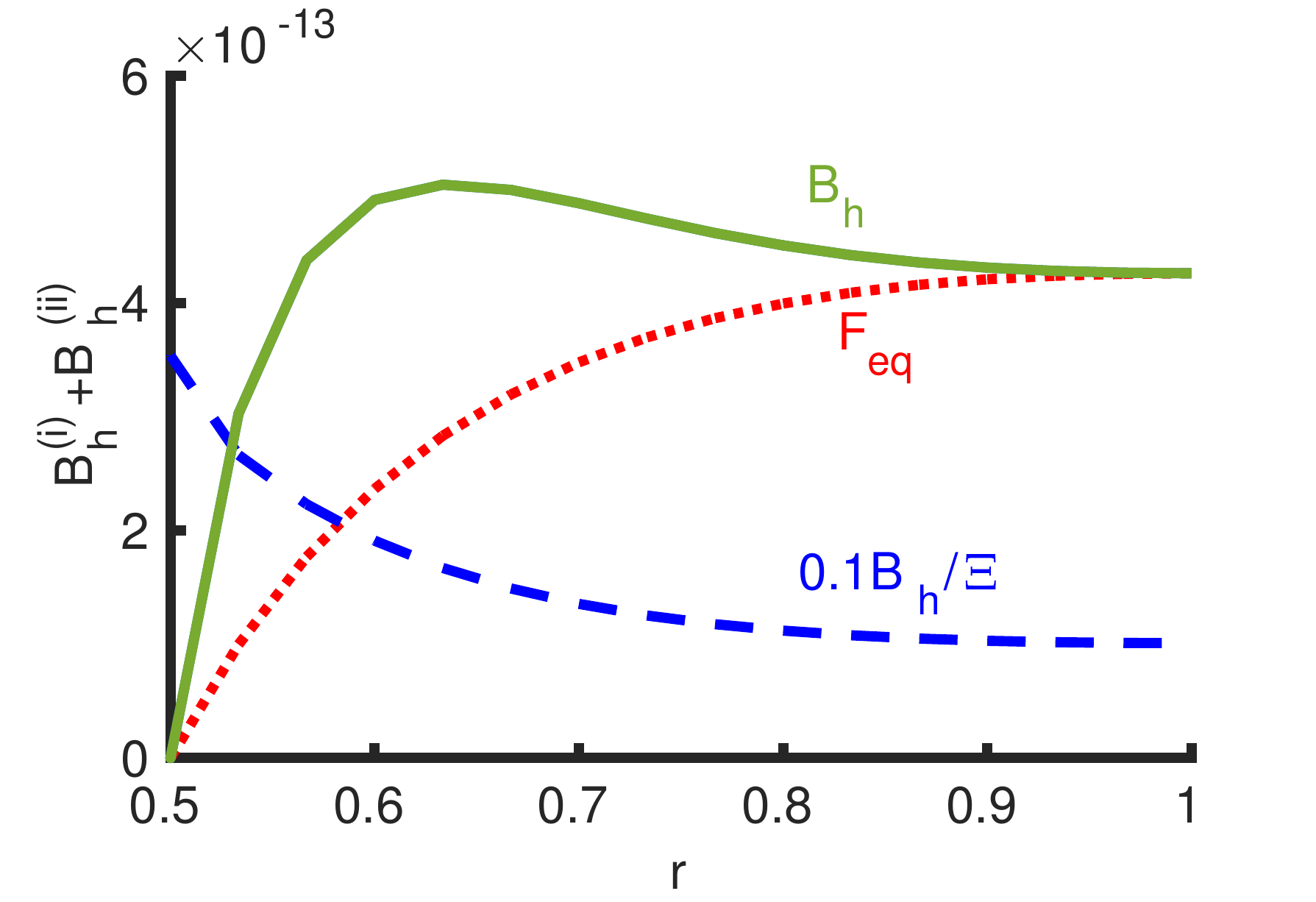}
\end{center}
  \caption
 { $(k_1k_2k_3)^2(B_h^{(i)}+B_h^{(ii)})$ is plotted as solid green line
 as an function of $r\equiv r_2=r_3$. The peak is located at $r\approx 0.6$ and the deviation from the equilateral shape (red dotted) is remarkable. 
The blue dashed line shows the case without $\Xi$, namely $10^{-1}(k_1k_2k_3)^2(B_h^{(i)}+B_h^{(ii)})/\Xi$
which is multiplied by $0.1$ for illustrative purpose.
The parameters are given in eq.~\eqref{sample parameters}.}
  \label{r-Bh}
\end{figure}

The total contribution to the bispectrum from $\mathcal{N}_i$ and
$\tilde{\mathcal{N}}_i$ is maximal in the folded limit
$r=0.5$. However, $\Xi(r_2,r_3)$ arising from the tensor polarisations
is also an important factor determining the shape of the tensor bispectrum. 
$\Xi$ is multiplied to the total bispectrum eq.~\eqref{Total Bh}
as an overall factor and it vanishes at $r_2=r_3=0.5$. In figure~\ref{r-Bh}, we illustrate
how $\Xi$ changes the shape of the bispectrum on the $r_2=r_3$ plane. $\Xi$ suppresses the bispectrum at lower $r$ and vanishes at $r=0.5$. In fact, $\Xi$ vanishes not only at $r_2=r_3=0.5$, but at all points on the line $r_2+r_3=1$ (i.e. the folded limit), because of conservation of angular momentum. The Feynman diagrams in figure~\ref{fig:Feynman}  can be seen as processes in which two spin-2 particles collide and one spin-2 particle comes out.
In particular, in  the case of a head-on collision, which corresponds to the folded limit, $k_2+k_3=k_1$, the cross-section vanishes, because the angular momentum is contributed only by spins (i.e. no orbital angular momentum) and the spin of the system cannot be conserved as $2\pm 2\neq\pm2$.
%
%

In summary, the peak of the tensor bispectrum $B_h$ is located
at $r\equiv r_2=r_3\approx 0.6$ for the following two reasons -
(i) Among several contributions to the sourced tensor bispectrum $B_h^{RRR}$, the biggest one comes from $\langle \hat{\psi}^R_2(k_1) \hat{\psi}^R_1(k_2)\hat{\psi}^R_1(k_3)\rangle\propto \mathcal{N}_1$ in which the two linear perturbations of $SU(2)$ gauge field $\hat{t}^R_1(k=rk_1)$
non-linearly induce the second order one $\hat{t}_2^R(k_1)$ and subsequently  $\hat{t}_2^R(k_1)$ sources the second order GW $\hat{\psi}^R_2(k_1)$.
In this process, the amplitude of the second order fluctuations is larger
when the momentum of the first order perturbations, $rk_1$, is smaller,
because in this case $t_1^{R}$ gets amplified when the second order $t_2^R$
is still deep inside the horizon where the source effect is more efficient
(i.e. Green's function $G_t$ has a bigger amplitude). 
Hence, $\mathcal{N}_1$, which dominates the tensor bispectrum, is a decreasing function of $r$ (see figure~\ref{Nir}).
(ii) The polarisation tensors also yield a $r$ dependence
as an overall factor $\Xi(r)$ to the total bispectrum.
$\Xi(r)$ is a growing function of $r$ and vanishes at $r=0.5$. Multiplying $\Xi(r)$ changes the blue dashed line into the green line in figure~\ref{r-Bh}.
As the result of (i) and (ii), we obtain the bispectrum with a peak at $r\simeq 0.6$.

\section{Parameter Search}
\label{sec:params}
In this section, we constrain the parameter regions from present observations and self-consistency of the model. We also clarify the parameter regions
where the power spectrum or bispectrum of the sourced GWs will be detectable by upcoming CMB observations.
Note that there remain four parameters, $H, m_Q, \epsilon_B$ and $g$,
and one relationship $g^2\epsilon_B\Mpl^2 = m_Q^4 H^2$. Eliminating $g$,
we are left with three free parameters, $H, m_Q$ and $\epsilon_B$, in our model.

\subsection{Tensor-to-scalar ratio}
Currently the CMB observations put an upper bound on the tensor-to-scalar ratio $r$ as
\begin{equation}\label{eq:planck_r}
r\equiv \frac{P_h(k_{\rm CMB})}{P_\zeta(k_{\rm CMB})}<0.07,
\qquad
({\rm 95\%\ C.L.}),
\end{equation}
where $P_{\zeta}$ is the power spectrum of the curvature perturbation
and $k_{\rm CMB}=0.05\,{\rm Mpc}^{-1}$.
In our model, not only the vacuum fluctuation of the metric but also
the sourced GWs contribute to $P_h$. 
Substituting eq.~\eqref{eq:pk_hh}, the total tensor-to-scalar ratio  is given as
\begin{equation}
r = \frac{\Delta^h_{\rm vac}}{\Delta_{\zeta}}\left(1+\frac{\epsilon_B}{2} \left|\mathcal{F}(m_Q)\right|^2\right)\,,
\end{equation}
where the dimensionless scalar power spectrum, $\Delta_{\zeta}\equiv
k^3P_{\zeta}/2\pi^2\approx 2.2\times 10^{-9}$~\cite{ade:2015}, and that
of the tensor metric vacuum fluctuation,
$\Delta^h_{\text{vac}} \equiv k^3 P_h^{\rm vac}/2\pi^2= 2H^2/\pi^2M^2_P$, are introduced. Translating the upper bound on $r$ into the constraint on our model parameters, we obtain
\begin{equation}\label{eq:eb_r}
\epsilon_B < \Big(\frac{0.07}{r_{\rm vac}}-1\Big)\frac{2}{|\mathcal{F}(m_Q)|^2}\,,
\end{equation}
where the conventional tensor-to-scalar ratio contributed only from the
tensor metric vacuum fluctuation is defined by
\begin{equation}
r_{\rm vac} \equiv \frac{\Delta^h_{\rm vac}}{\Delta_{\zeta}}\approx
\frac{1}{2.2\times 10^{-9}}\frac{2H^2}{\pi^2 \Mpl^2}. 
\end{equation}

Since the upcoming CMB B-mode polarisation observation missions aim to achieve
a sensitivity $r\approx 10^{3}$, the parameter region predicting $r\ge 10^{-3}$
is particularly interesting. In our model, we find
\begin{equation}
r\ge10^{-3} \quad\Longleftrightarrow\quad
\epsilon_B \geq \Big(\frac{10^{-3}}{r_{\rm vac}}-1\Big)\frac{2}{|\mathcal{F}(m_Q)|^2}\,.
\end{equation}

\subsection{Tensor bispectrum}
The constraint on the tensor bispectrum in the equilateral limit is also reported as a bound on $f_{\rm NL}^{\rm tens}$~\cite{ade:2015};\footnote{The factor of $2\sqrt{2}$ in the denominator comes from the difference of the normalisation of the polarisation tensors. In \cite{ade:2015}, 
$e^R_{ij}(\bm{k})e^R_{ij}(-\bm{k})=2$ is adopted. }
\begin{equation}
-1100<f_{\rm NL}^{\rm tens} \equiv \frac{B_h^{RRR}(k,k,k)}{2\sqrt{2}F^{\rm eq}_{\zeta}(k)}
<1900\,,
\qquad ({\rm 68\%\ C.L.}),
\end{equation}
where $B^{RRR}_h(k_1,k_2,k_3)$ is defined in eq.~\eqref{BRRR def} and $F^{\rm eq}_{\zeta}(k) \equiv (18/5)P^2_{\zeta}(k)$, evaluated at the pivot scale, $k_{\text{CMB}} = 0.05\, \text{Mpc}^{-1}$~\cite{ade:2015}.
Our model should satisfy these two observational constraints.

From the constraint on $f^{\text{tens}}_{\text{NL}}$ we find,
\begin{align}\label{eq:eb_fnl}
\epsilon_B &\geq \frac{64\, M^4_{\text{Pl}}}{27\, H^4}\frac{1100\cdot 18\cdot \Delta^2_{\zeta}\cdot 4\pi^4\cdot 2\sqrt{2}}{-5\cdot\Upsilon_{\rm eq}(m_Q)}\,,\\
\epsilon_B &\leq \frac{64\, M^4_{\text{Pl}}}{27\, H^4}\frac{1900\cdot 18\cdot \Delta^2_{\zeta}\cdot 4\pi^4\cdot 2\sqrt{2}}{5\cdot\Upsilon_{\rm eq}(m_Q)}\,,
\end{align}
where the first constraint applies when $\Upsilon_{\rm eq}<0$ and the second when $\Upsilon_{\rm eq}>0$. 

\subsection{Consistency of the model}
In addition to these observational constraints, we discuss the restriction imposed by self-consistency of the model.
Scalar perturbations of the spectator sector
have a fatal instability on sub-horizon scale if $m_Q<\sqrt{2}$~\cite{Dimastrogiovanni:2012ew}. Hence we demand $m_Q>\sqrt{2}$ in our model.
Since $\epsilon_B$ approximately indicates the energy density fraction of the background SU(2) gauge field,
\begin{equation}
\Omega_A\equiv \frac{\rho_Q}{3\Mpl^2H^2}= \frac{(\dot{Q}+HQ)^2+g^2Q^4}{2\Mpl^2H^2}
\simeq \frac{1+m_Q^{2}}{2m_Q^2}\,\epsilon_B, 
\end{equation}
$\epsilon_B$ is positive and small.
As found in~\cite{Fujita:2017jwq}, if $\epsilon_B$ is too large, its effect on the evolution of the inflaton perturbation significantly alters the spectral index $n_s$, because $\epsilon_B$ contributes to $\dot{H}$ through eq.~\eqref{Friedmann2}.
To keep this effect negligible, it is  required  
\begin{align}\label{eq:slow_roll}
\epsilon_B(t_{*}) < 2\times 10^{-2}\,,
\end{align}
where $t_{*}$ is the time at which CMB modes leave the horizon. 
On the other hand, since $\epsilon_B$ can be rewritten as $\epsilon_B= m_Q^4 H^2/(g^2\Mpl^2)$, if one lowers $\epsilon_B$ by fixing $m_Q$ and $H$,
one would confront a large self-coupling constant $g$ of the SU(2) gauge fields which leads to a non-negligible backreaction from $SU(2)$ tensor perturbations
to the background dynamics.
In order to avoid large backreaction, we need to have~\cite{dimastrogiovanni/fasiello/fujita:2016}
\begin{equation}\label{eq:eb_br}
\epsilon_B > \frac{m^2_Q[\mathcal{B}+\tilde{\mathcal{B}}/(m_Q+m^{-1}_Q)]}{24\pi^2} \Big(\frac{H}{\Mpl}\Big)^2\,, 
\end{equation}
where $\mathcal{B}$ and $\tilde{\mathcal{B}}$ are functions of $m_Q$ given by
\begin{align}
\mathcal{B}(m_Q) = \int_0^{x_{\text{max}}} dx\, x \left| i^{\beta} W_{\beta, \alpha}(-2ix)\right| ^2\,,\quad  
\tilde{\mathcal{B}}(m_Q) = \int_0^{x_{\text{max}}} dx\, x^2 \left| i^{\beta} W_{\beta, \alpha}(-2ix)\right| ^2.
\end{align}

We also find that \refeq{eb_br} ensures $g \ll 1$ as well, which is preferred for validity of the perturbation series.\footnote{Strictly speaking, since our setup does not include any $SU(2)$ charged particle, a large $g$ itself is not necessarily problematic. However, if one considers a charged particle, $g\gtrsim 1$ causes a strong coupling problem in that loop effects would alter dynamics of the SU(2) gauge fields.}

\subsection{Allowed parameter regions}
\begin{figure*}
        \centering
        \includegraphics[width=1\textwidth]{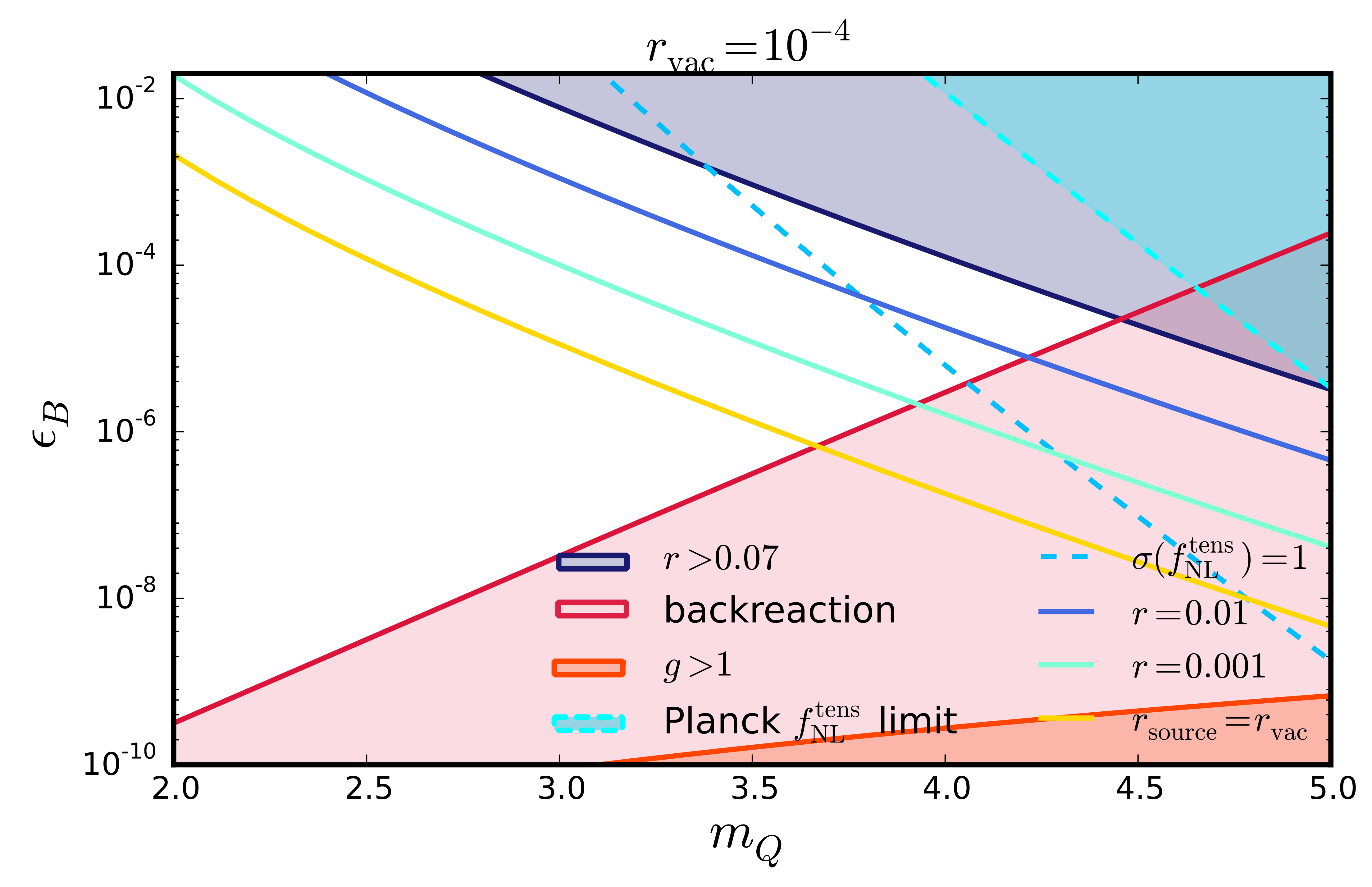}
 \caption{ 
 Parameter space for gravitational wave production in our model, for $r_{\rm vac} = 10^{-4}$, and for scale-invariant GWs. The blue and magenta shaded regions are excluded by 
the current upper bound on $f_{\rm NL}^{\rm tens}$ and
 $r$~\cite{ade:2015}, respectively.
The light red shaded region is not necessarily ruled out but a significant
backreaction requires a dedicated numerical treatment to obtain the predictions. In the orange shaded region, the system confronts  a strong coupling problem, if one considers $SU(2)$ charged particle. We also show $f^{\rm tens}_{\rm NL}=1$ as the dashed blue line, because an error of order unity $\sigma(f^{\text{tens}}_{\text{NL}})\sim 1$ would be achieved by upcoming CMB B-mode missions. The solid lines denote $r=10^{-2}$ (blue), $10^{-3}$ (green) and $10^{-4}$ (yellow).}
        \label{fig:param_1}
\end{figure*}

%
\refEqs{planck_r}{eb_br}, together with $m_Q>\sqrt{2}$, give the set of constraints we employ to define the regions in $\epsilon_B$-$m_Q$ plane that are interesting for future CMB experiments. \refFigs{param_1}{param_3} show the allowed regions for 3 different choices of $H$ or equivalently $r_{\rm vac} = 10^{-4}$, $10^{-3}$ and $10^{-2}$. As $r_{\rm vac}$ increases, the allowed parameter space shrinks. This is because the upper bound on $r$ implies that the power in the sourced tensor modes cannot be very large if $r_{\rm vac}$ is large.

The bottom right corners in these figures (i.e. regions with a large $m_Q$ and small $\epsilon_B$) are shaded as the parameter spaces with non-negligible backreaction, although this does not mean that these regions are excluded.
Rather, it indicates that one needs to perform numerical calculations to
take into account backreaction, to study this parameter space~\cite{dimastrogiovanni/fasiello/fujita:2016}
(see ref.~\cite{Fujita:2017jwq} where the backreaction is numerically incorporated). 

We find that there's a general trend in the constraining power of tensor
power spectrum and bispectrum. While the power spectrum is better at
constraining small $m_Q$ regions, the bispectrum is better at
constraining large $m_Q$ regions. This happens because $B_h$ is
exponentially more sensitive to $m_Q$ (it has an extra factor of
$e^{2\pi(2m_Q+m^{-1}_Q)}$ compared to $r$), and so, a small change in
$m_Q$ can easily change $B_h$ by a large factor ($\sim e^{4\pi \Delta
m_Q}$). This also has interesting consequences for detectability of the
tensor bispectrum, as a large range of bispectra can be generated even
for the small range of values of interest, $\sqrt{2} < m_Q \apprle 4$,
making a detection of the tensor bispectrum (in this model) possible in
the near future, even if $r$ is small (see
\reffigs{param_1}{param_3}). While even the current constraints on
$f^{\text{tens}}_{\rm NL}$ are useful for ruling out the top right
corners in the figures (i.e. regions with a large $m_Q$ and large
$\epsilon_B$), there remains parameter space in which the tensor bispectrum can be observed in the future. It is then natural to ask what range of parameters can be probed in upcoming CMB missions.  

\begin{figure*}
        \centering
        \includegraphics[width=1\textwidth]{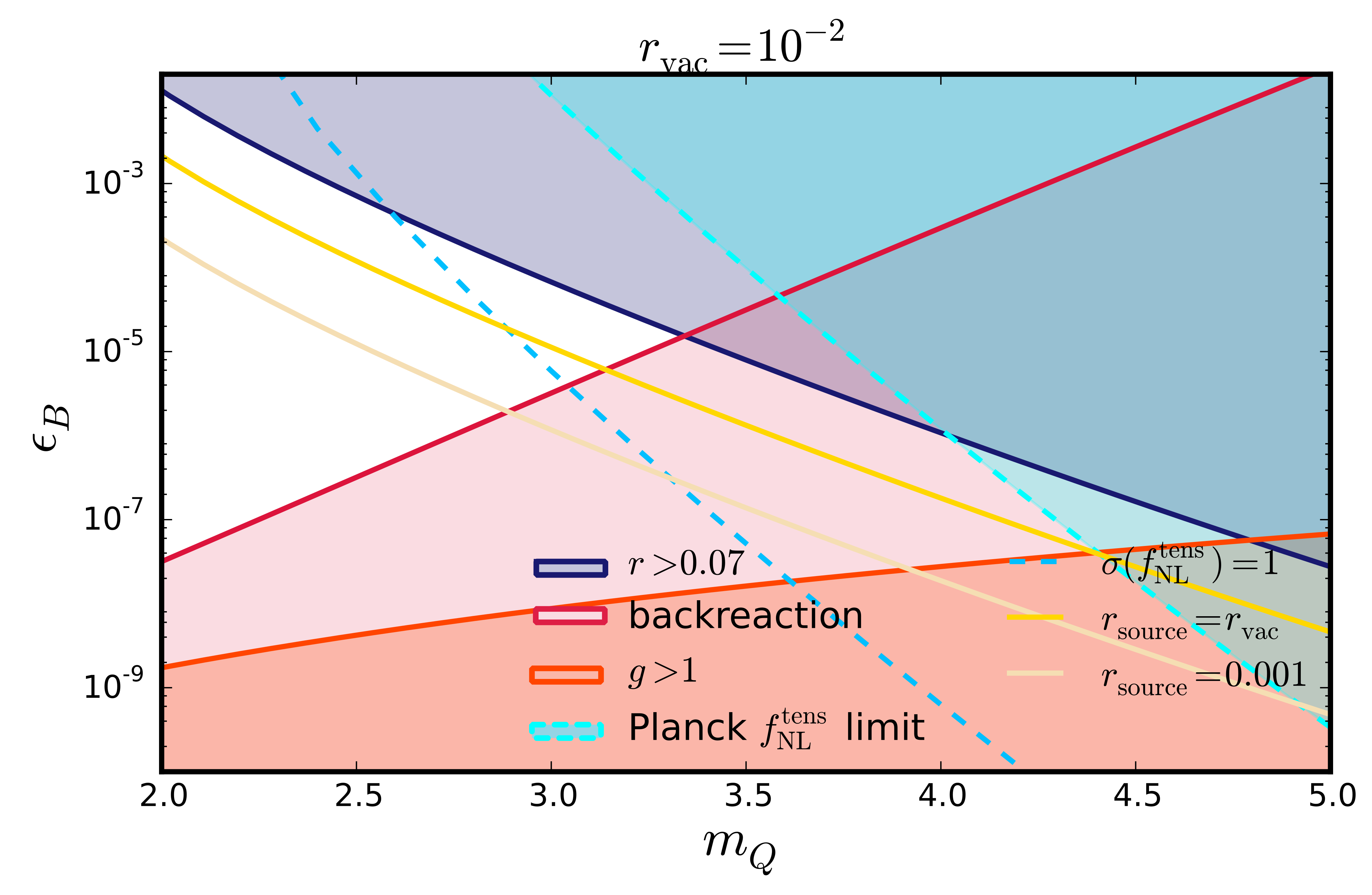}
        \caption{Same as \reffig{param_1} but for $r_{\rm vac} = 10^{-2}$.}
        \label{fig:param_3}
\end{figure*}

To that end we also plot the line for
$\sigma(f_{\text{NL}}^{\text{tens}})=1$ in \reffigs{param_1}{param_3},
which is expected to be the target sensitivity of LiteBIRD
(M. Shiraishi, private communication). We see that this improved
sensitivity will allow us to probe a significant portion of the
parameter space with large $m_Q$ and small $\epsilon_B$, which is
inaccessible to measurements of $r$, even if we can measure
$r=10^{-4}$~(\reffig{param_1}, bottom right). Although our present
calculation does not ensure that this conclusion stays unchanged when we
account for the backreaction, it might still be true when backreaction
is included. We also show the line corresponding to $r_{\text{source}} =
r_{\text{vac}}$. Regions to the left of this line denote regions where
the amplitude of the sourced tensor modes is smaller than the amplitude
of vacuum tensor of the metric. From \reffig{param_3} we see that if
$r_{\text{vac}} = 10^{-2}$, there is a region of intermediate $\epsilon_B$ and $m_Q$ values for which $f^{\text{tens}}_{\text{NL}} > 1$, also if $r_{\text{source}} < r_{\text{vac}}$.  This regime is particularly interesting because one can learn about both vacuum fluctuations of the metric and spectator fields during inflation, by combining the power spectrum and bispectrum. On the other hand, if $r_{\text{vac}}$ is smaller, a small $r_{\text{source}}$ is accompanied by a small tensor bispectrum as well.

\begin{figure*}
        \centering
        \includegraphics[width=1\textwidth]{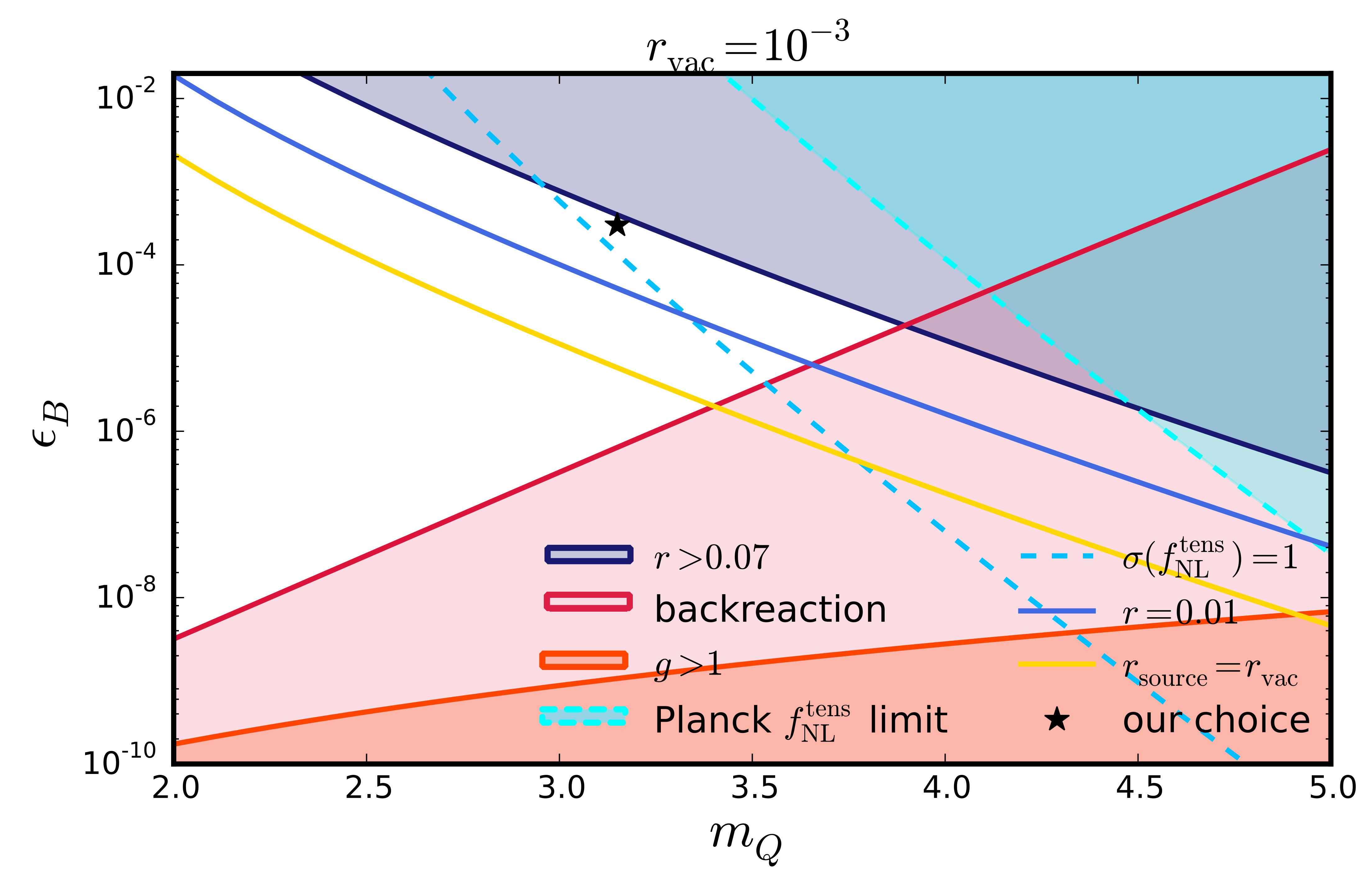}
        \caption{Same as \reffig{param_1} but for $r_{\rm vac} = 10^{-3}$.
        The black star denotes the parameter choice given in eq.~\eqref{sample parameters}. }
        \label{fig:param_2}
\end{figure*}

\section{Conclusion}
\label{sec:conclusion}
%
%
%

In this paper we have calculated the bispectrum of tensor perturbations
sourced by spectator SU(2) gauge fields during
inflation~\cite{dimastrogiovanni/fasiello/fujita:2016}.
The primary contribution to the bispectrum comes from the
self-interaction of the SU(2) gauge fields; thus, it is unique to
non-Abelian gauge theory. We find that the amplitude of the bispectrum
parametrised by its ratio to the (squared) power spectrum, $B_h/P^2_h$,
is very large, $\sim 1/\epsilon_B$ \cite{Agrawal:2017awz}. Since
$\epsilon_B \ll 1$, this is much larger than $\sim 1$ which is predicted
for quantum fluctuations of the metric \cite{maldacena:2002,maldacena/pimentel:2011}.

We also explored parameter space of the model relevant to future CMB
missions. Even with an $r_{\rm vac}$ as low as $10^{-4}$, large
parameter space remains consistent theoretically as well as with the
current CMB observations. However, the exponential sensitivity of the power
spectrum and bispectrum on model parameters makes it difficult to
completely eliminate all the parameter space of the model
on the basis of just these observations.

%
%

Upcoming CMB missions such as LiteBIRD~\cite{matsumura/etal:2013} will
measure the CMB polarisation to unprecedented accuracy. This will allow
us to not only detect B-modes but also to characterise them, hence
testing one of our most ambitious claims about our origins. If the
primordial B-modes arise from quantum fluctuations of the metric, we
will find them to be parity invariant, near scale-invariant, and weakly
non-Gaussian. If not, we will use the deviations to constrain the
fraction of energy density in spectator gauge fields in the inflationary
Universe~\cite{cook/sorbo:2013,namba/etal:2015,shiraishi/etal:2016,Agrawal:2017awz}.

\acknowledgments
We would like to thank M. Shiraishi and R. Namba for useful discussions. 
\appendix

\section{Polarisation Tensor}
\label{Polarisation Tensor}

In this appendix, we construct the left and right-handed transverse and traceless polarisation tensors.
We start with the left and right-handed polarisation vector whose wave number is parallel to the z-axis, 
\begin{equation}
\bm{\epsilon}^{L/R}(\hat{\bm z})= \frac{1}{\sqrt{2}}
\begin{pmatrix}1 \\
\pm i \\
0 \\
\end{pmatrix}.
\end{equation}
The plus and minus signs are for left- ($L$) and right-handed
($R$) polarisation vectors, respectively. From now on, $\pm$ means $+$ for
$L$ and $-$ for $R$, whereas $\mp$ means $-$ for $L$ and $+$ for $R$. To obtain the polarisation vector with a general wave number $\hat{\bm k}$
which points in the direction of $(\theta,\varphi)$ in polar coordinate,
we use the following rotation matrix which transforms $\hat{\bm z}$ into $\hat{\bm k}$:
\begin{equation}
S(\hat{\bm k})=\begin{pmatrix}
\cos\theta \cos\varphi\ & -\sin\varphi\ & \sin\theta\cos\varphi \\
\cos\theta\sin\varphi & \cos\varphi & \sin\theta\sin\varphi \\
-\sin\theta & 0 & \cos\theta \\
\end{pmatrix}.
\end{equation}
Then we find
\begin{equation}
\bm{\epsilon}^{L/R}(\hat{\bm k})=S(\hat{\bm k}) \bm{\epsilon}^{L/R}(\hat{\bm z})= \frac{1}{\sqrt{2}}
\begin{pmatrix}
\cos\theta\cos\varphi \mp i \sin\varphi \\
\cos\theta\sin\varphi\pm i \cos\varphi \\
-\sin\theta \\
\end{pmatrix}.
\end{equation}
These polarisation vectors satisfy
\begin{align}
&\bm{k}\cdot \bm{\epsilon}^{L/R}(\hat{\bm k})=0,
\qquad
\bm{\epsilon}^{L/R*}(\hat{\bm k})=\bm{\epsilon}^{R/L}(\hat{\bm k})
=\bm{\epsilon}^{L/R}(-\hat{\bm k}),\notag\\
&
\bm{\epsilon}^{L/R}(\hat{\bm k})\cdot\bm{\epsilon}^{R/L}(\hat{\bm k})=1,
\qquad
\bm{\epsilon}^{L/R}(\hat{\bm k})\cdot\bm{\epsilon}^{L/R}(\hat{\bm k})=0.
\label{pvector property}
\end{align}
The polarisation tensor $e^{L/R}_{ij}(\hat{\bm k})$ can be constructed from the polarisation vectors,
\begin{equation}
e^{L/R}_{ij}(\hat{\bm k})
= \epsilon_i^{L/R}(\hat{\bm k})\, \epsilon_j^{L/R}(\hat{\bm k}).
\end{equation}
These polarisation tensors are transverse and traceless and satisfy
\begin{equation}
e^{L}_{ij} (-\hat{\bm{k}})= e^{L*}_{ij} (\hat{\bm{k}})=e^{R}_{ij} (\hat{\bm{k}}),
\quad
i \epsilon_{ijk} k_i e_{jl}^{L/R}(\hat{\bm{k}})=
\pm k e_{kl}^{L/R}(\hat{\bm{k}}),
\quad
e^{L}_{ij} (\hat{\bm{z}}) =\frac{1}{2}\left(\begin{array}{ccc}
1 & i & 0 \\
i & -1 & 0 \\
0 & 0 & 0 \\
\end{array}\right).
\label{Pol property}
\end{equation}

Although the general expression for $e^{L/R}_{ij}(\hat{\bm k})$ is rather complicated, we can fix $\theta$ in the current case. This is because 
we calculate three polarisation tensors with three different wavenumbers,
$e_{ij}^R(\hat{\bm k}_1) e_{kl}^R(\hat{\bm k}_2) e_{nm}^R(\hat{\bm k}_3)$
whose indices are somehow contracted, and these wave vectors are on the same plane due to momentum conservation, $\delta(\bm k_1+\bm k_2 +\bm k_3)$.
In that case, we can set $\theta = \pi/2$ and let these vectors, $\bm k_1, \bm k_2, \bm k_3$, move only on the x-y plane. For $\theta=\pi/2$,
the polarisation tensors become
\begin{equation}
e_{ij}^{L/R}\left(\theta=\frac{\pi}{2},\varphi\right)=\frac{1}{2}
\begin{pmatrix}
-\sin^2\varphi\ & \cos\varphi \sin \varphi \ & \pm i \sin\varphi \\
\cos\varphi\sin\varphi& -\cos^2\varphi & \mp i \cos\varphi \\
\pm i \sin\varphi\ & \mp i \cos\varphi & 1 \\
\end{pmatrix}.
\end{equation}
Now we have three angles, $\varphi_1, \varphi_2, \varphi_3$, associated with wavenumbers, $\bm k_1, \bm k_2, \bm k_3$, respectively. Without loss of generality, we can set $\varphi_1=0$.
Furthermore, these trigonometric functions of $\varphi_2$ and $\varphi_3$ can be rewritten as functions of $r_2\equiv k_2/k_1$ and $r_3\equiv k_3/k_1$. Using $\bm k_1+\bm k_2 +\bm k_3=0$, we find
\begin{equation}
k_3^2=|\bm k_1+\bm k_2|^2= k_1^2+k_2^2+2 k_1 k_2\cos\varphi_2
\quad \Longrightarrow\quad
\cos\varphi_2 = \frac{r_3^2-r_2^2-1}{2r_2}.
\end{equation}
In the same way, we also find $\cos\varphi_3=(r_2^2-r_3^2-1)/(2r_3)$.
With this notation, we find
\begin{align}
&\sum_{\{I,J,K\}}^{\{1,2,3\}} i \epsilon^{abc}\, 
k_{K}^i e_{ai}^R(\hat{\bm{k}}_I)e_{bj}^R(\hat{\bm{k}}_J)e_{cj}^R(\hat{\bm{k}}_K)
=-2k_1  \Xi\, \tilde{\Xi}\,, 
\\
&e_{ij}^R(\hat{\bm{k}}_1)e_{jl}^R(\hat{\bm{k}}_2)e_{li}^R(\hat{\bm{k}}_3)
=\Xi,\qquad
\epsilon^{abc} \epsilon^{ijk} \, e_{ai}^R(\hat{\bm{k}}_1)e_{bj}^R(\hat{\bm{k}}_2)e_{ck}^R(\hat{\bm{k}}_3)
=2\,\Xi\,,
\label{pol sum}
\end{align}
where  
\begin{align}
\Xi
&\equiv \frac{(1+r_2+r_3)^3}{64r_2^2r_3^2} (r_2+r_3-1)(1+r_2-r_3)(1+r_3-r_2)\,,
\\
\tilde{\Xi}&\equiv1+r_2+r_3.
\end{align}
$\sum_{\{I,J,K\}}^{\{1,2,3\}}$ denotes summation of all the permutation, $\{ I,J,K\}=\{{\rm Perm}(1,2,3)\}$.

\section{Equilateral Shape}
\label{Equilateral Shape}

\begin{figure*}
        \centering
        \includegraphics[width=1\textwidth]{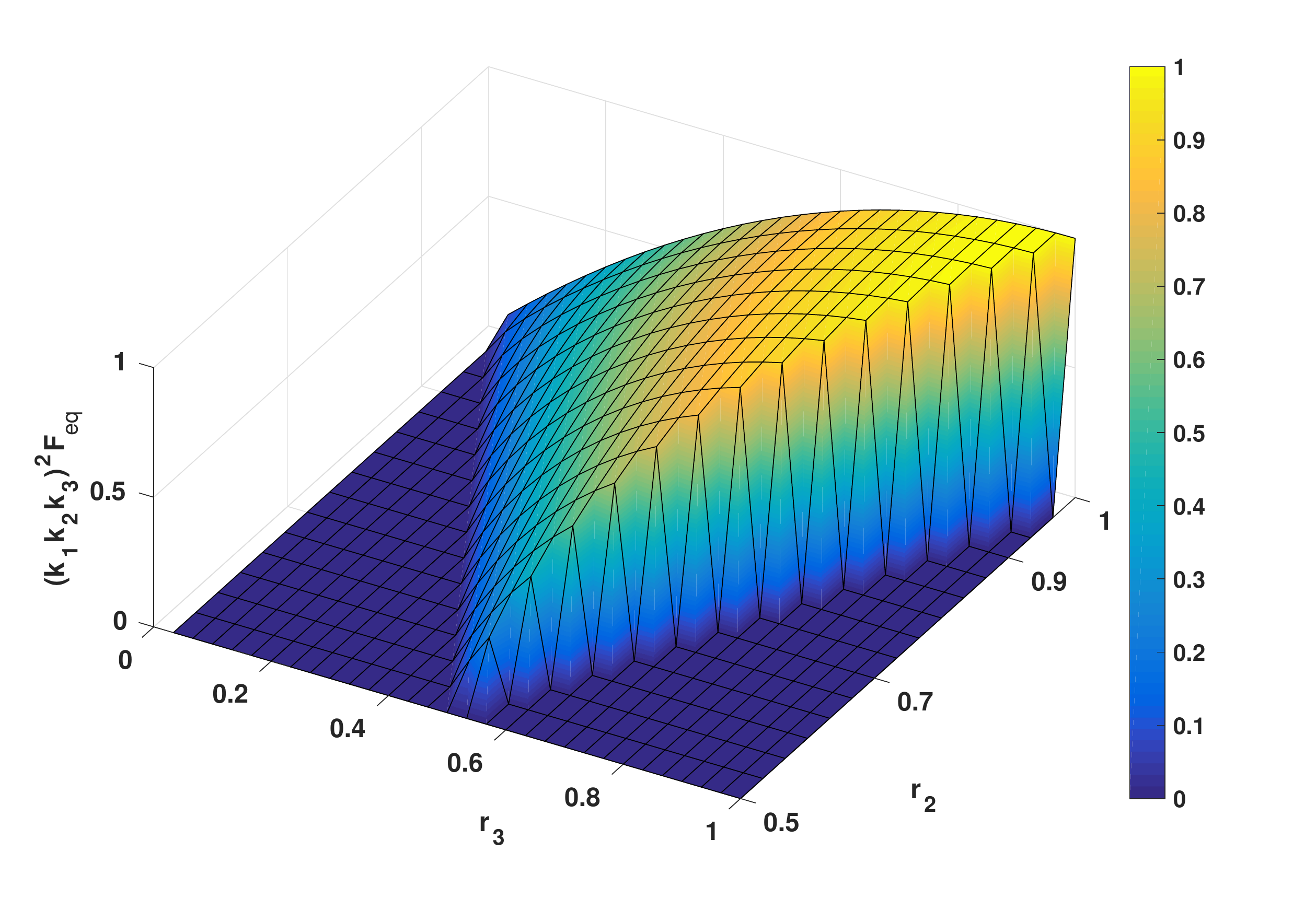}
        \caption{3D plot of $(k_1k_2k_3)^2F_{\text{eq}}$, \refeq{eq_shape}.}
        \label{fig:eq_shape}
\end{figure*}

To measure similarity of the shapes of bispectra, the cosine
between two shapes is introduced as~\cite{babich/creminelli/zaldarriaga:2004},
\begin{equation}
\cos (B_h,F_{\text{ref}})\equiv \frac{B_h \cdot F_{\text{ref}}}{\sqrt{(B_h \cdot B_h)(F_{\text{ref}}\cdot F_{\text{ref}})}}\,,
\end{equation}
where the dot product is defined as 
\begin{equation}
X \cdot Y \equiv \int_0^1\, dr_2 \, \int_0^1\, dr_3 (r_2r_3)^4 X(1, r_2, r_3)Y(1, r_2, r_3).
\end{equation}
Here $F_{\text{ref}}$ is the reference template to which the similarity is measured.
In this paper we use the equilateral template~\cite{creminelli/etal:2006} \begin{equation}\label{eq:eq_shape}
F_{\text{eq}}(k_1, k_2, k_3) = \Bigg[-\frac{1}{k^3_1k^3_2}-\frac{1}{k^3_1k^3_3}-\frac{1}{k^3_2k^3_3}-\frac{2}{k^2_1k^2_2k^2_3}+\frac{1}{k_1k^2_2k^2_3}+(5\, \text{perm})\Bigg]\,.
\end{equation}
\refFig{eq_shape} shows the shape of this template as a function of $r_2$ and $r_3$. 
\bibliography{references}

\end{document}